\documentclass[preprint2]{proto}
\usepackage{times}
\usepackage{amsmath}
\usepackage[T1]{fontenc}

\usepackage{color, colortbl}
\usepackage[utf8]{inputenc} 
\usepackage[T1]{fontenc}    
\usepackage{url}            
\usepackage{booktabs}       
\usepackage{amsfonts}       
\usepackage{nicefrac}       
\usepackage{microtype}      
\usepackage{lipsum}
\usepackage{array}

\definecolor{Gray}{gray}{0.9}

\begin{document}

\title{\textbf{\LARGE Trans-Neptunian Space and the Post-Pluto Paradigm}}

\author{
 \textbf{\large  Alex H. Parker}
}

\affil{\small\em Department of Space Studies\\
  Southwest Research Institute\\
  Boulder, CO 80302}

\begin{abstract}
\begin{list}{ } {\rightmargin 1in}
\baselineskip = 11pt
\parindent=1pc
{\small 

The Pluto system is an archetype for the multitude of icy dwarf planets and accompanying satellite systems that populate the vast volume of the solar system beyond Neptune. New Horizons' exploration of Pluto and its five moons gave us a glimpse into the range of properties that their kin may host. Furthermore, the surfaces of Pluto and Charon record eons of bombardment by small trans-Neptunian objects, and by treating them as witness plates we can infer a few key properties of the trans-Neptunian population at sizes far below current direct-detection limits. This chapter summarizes what we have learned from the Pluto system about the origins and properties of the trans-Neptunian populations, the processes that have acted upon those members over the age of the solar system, and the processes likely to remain active today. Included in this summary is an inference of the properties of the size distribution of small trans-Neptunian objects and estimates on the fraction of binary systems present at small sizes. Further, this chapter compares the extant properties of the satellites of trans-Neptunian dwarf planets and their implications for the processes of satellite formation and the early evolution of planetesimals in the outer solar system. Finally, this chapter concludes with a discussion of near-term theoretical, observational, and laboratory efforts that can further ground our understanding of the Pluto system and how its properties can guide future exploration of trans-Neptunian space.

 \hfill \break
 }
\end{list}
\end{abstract}

\section{\textbf{INTRODUCTION}}

Eighty-five years of thought regarding the potential properties of Pluto preceded our first exploration of it and its satellites with the 2015 New Horizons flyby. In that time, theories draped around what could be observed remotely produced a long series of predictions that were tested at flyby. Many held up, while others did not. 

Fewer than 20 years have elapsed since the first discoveries of worlds beyond Neptune that can rightfully be called Pluto's kin. Our understanding of these worlds -- Eris, Makemake, Haumea, and others -- started from an advantaged position, launching as it did from the existing understanding that had developed regarding the Pluto system. However, a twist of fate has placed these three most similar trans-Neptunian dwarf planets at or near their aphelia in the current epoch. Thus, these worlds are universally dimmer and more remote than Pluto and in the time we have had to consider them we have struggled to build as compelling a body of observation and theory as existed for Pluto prior to flyby. Their individual uniquenesses are notable but poorly understood.

As new facilities and instruments are developed, this will all change. The near future will deliver a multitude of opportunities to build an observational understanding of the surfaces of the largest trans-Neptunian objects (TNOs) and their environments that rivals or exceeds that which we had available at the Pluto flyby. Perhaps the greatest advance, however, will come from taking stock of what we now know to be possible about these worlds thanks to the vast wealth of in-situ information delivered by New Horizons' exploration of the Pluto system. The largest TNOs will all have properties that make them unique from one another, but Pluto will long serve as a proving ground for testing new ideas about the general properties of these distant, complex worlds, and the processes that shape them. 

To understand both the promise and limits of the Pluto system as an archetype of the distant dwarf planets that fill the void beyond Neptune, it is first important to understand its relationships to the numerous populations that call this region home; including their respective origins and evolution, and the extent of their present interactions.

\bigskip
\noindent
\textbf{1.1 The Structure Of Trans-Neptunian Space}
\bigskip

Trans-Neptunian space is occupied by a host of sub-populations of small bodies, distinguished by both dynamical and physical properties. These sub-populations and their dynamical properties have characterized by a relatively small number of largely ground-based, wide-field observational programs conducted since the late 1990s. The Deep Ecliptic Survey (DES; Millis et al. 2002, Buie et al. 2003, Elliot et al. 2005, Gulbis et al. 2010, Adams et al. 2014) provided the first broad-brush well-characterized look at the dynamical divisions in trans-Neptunian populations. The Canada-France Ecliptic Plane Survey (CFEPS) used the 3.5m Canada-France-Hawaii Telescope and its large-format MegaCam imager to conduct an extremely well-characterized deep and wide survey, leading to many of the current best-estimate measurements of the intrinsic orbital distributions of TNOs (Jones et al. 2006, Kavelaars et al. 2009, Petit et al. 2011, Gladman et al. 2012, Petit et al. 2017) and illustrating the necessity of careful survey design to account for discovery and tracking biases. The CFEPS effort led to a deeper follow-on program -- the Outer Solar System Origins Survey (OSSOS; Bannister et al. 2016a\&b, Shankman et al. 2016, Volk et al. 2016, Lawler et al. submitted) -- which was conducted on the same facility and for which analysis is still in progress. Ongoing efforts by several teams have begun to map the orbital distribution of the extremely distant Sedna-like population, which may hint at a large unseen perturber lurking at several hundred AU from the Sun (e.g., Trujillo \& Sheppard 2014, Batygin \& Brown 2016, Sheppard et al. 2019), though independent datasets to not all show evidence for the orbital distribution features upon which this hypothesis rests (e.g., Shankman et al. 2017). Together, these surveys have provided our best measurements of the intrinsic orbit and size distributions of the majority of TNO sub-populations.

Broadly, TNOs divide into sub-populations as follows. The scattered disk objects (SDOs) have perihelia near Neptune but bear no resonant protection against destabilizing close encounters. The resonant Kuiper Belt populations reside in numerous mean-motion resonances (MMRs) with Neptune including the highly populated 3:2, 5:2, and 2:1 resonances as well as many others to a lesser degree. The classical Kuiper Belt populations are in stable non-resonant orbits far from Neptune. The classical Kuiper Belt populations further divide into ``hot'' and ``cold'' sub-populations, where hot classical Kuiper Belt Objects (HCKBOs) have relatively excited inclinations and eccentricities, and the cold classical KBOs (CCKBOs) have very low-inclination, nearly-circular orbits that dominantly reside between 42 and 47 au. The CCKBO orbital distribution appears to have multiple components, with ``stirred'' and ``kernel'' sub-populations called out in some analyses (e.g., Petit et al. 2011). Finally, there is the relatively recently-recognized ``detached'' population (e.g., Trujillo \& Sheppard 2014, Sheppard et al. 2019), which have orbits that appear similar to SDOs except for their very high perihelia, which puts them well beyond the influence of any of the known giant planets. This population is exemplified by Sedna, and its origins and relationship with other populations remains a topic of debate. See Gladman et al. (2008) for a discussion of nomenclature.

The TNO luminosity function was measured in parallel with these efforts using deep ``pencil-beam'' surveys (Gladman et al. 1998, Gladman et al. 2001, Bernstein et al. 2004, Fuentes \& Holman 2008, Fraser \& Kavelaars 2009, Fuentes et al. 2009, Fraser et al. 2010, Fraser et al. 2014). These efforts traded dynamical resolution (i.e., short observational arcs versus long observational arcs) for survey depth, and produced estimates of the TNO luminosity function for ``hot'' and ``cold'' populations, defined by an inclination threshold usually around $5^{\circ}$. These surveys reached limiting absolute magnitudes approaching $H_{r'}\sim12$ mag, or roughly $D\sim17$ km for geometric albedos of $10\%$. Broadly, the luminosity function of any given sub-population is reasonably well characterized by a broken power-law of the form 

\begin{equation}
\frac{dN}{dH}=\begin{cases}
10^{\alpha_{1} (H-H_{0})} & H < H_{B} \\
10^{\alpha_{2} (H-H_{0}) + (\alpha_{1} - \alpha_{2})(H_B-H_{0})} & H \geq H_{0},
\end{cases}
\end{equation}

where $\alpha_{1}$ is the power-law slope valid for bright objects, $\alpha_{2}$ is the power-law slope valid for faint objects, $H_{B}$ is the absolute magnitude at which the slope transitions from $\alpha_1$ to $\alpha_2$, and $H_{0}$ is a normalization factor. The segments of this luminosity function each translate to a size-frequency distribution of the form $dN/dR \propto R^{-q}$ with a differential slope of $q=5\alpha+1$, where $\alpha$ is the local luminosity function slope. The complete form for a multiply-broken power law size-frequency distribution is further defined in Section 2. Canonically, a population in collisional equilibrium will reach a slope of $q=3.5$. TNOs with diameters larger than approximately 100 km have slopes much steeper than collisional equilibrium, while objects smaller than this size have slopes consistent or slightly shallower than collisional equilibrium (Bernstein et al. 2004, Petit et al. 2011, Fraser et al. 2014). 

Finally, the largest TNOs were discovered in wide field, relatively shallow surveys. These include the discoveries of Eris, Makemake, Haumea, Quaoar, Orcus, Sedna, and Gonggong (the recently-applied formal name for 2007 OR${10}$) at Palomar Observatory between 2002 and 2007 (Brown 2008). These largest TNOs are present in every TNO sub-population \emph{except} the low-inclination cold classical KBOs, where the size distribution truncates at an upper limit of a few hundred kilometers. 

\bigskip
\noindent
\textbf{1.2 Pluto's dynamical history and relationship to trans-Neptunian sub-populations}
\bigskip

The Pluto system orbits about the solar system barycenter in a multi-resonant configuration with Neptune, including the 3:2 MMR (Cohen \& Hubbard 1967) and a Kozai secular resonance (with the argument of perihelion librating around $90^\circ$; Milani et al. 1989). Currently, its mean inclination with respect to the solar system invariable plane is 15.55$^\circ$, near the peak of the inclination distributions of the other 3:2 resonators and other excited trans-Neptunian populations such as the Neptune Trojans, HCKBOs, and Scattered Disk Objects (SDOs). Because of the protection from Neptune close encounters conferred by its resonant configuration, the Pluto system remains stable even with its relatively extreme perihelion as compared to the majority of other trans-Neptunian populations, passing interior to Neptune at $q=29.66$ AU. 

Under all currently-debated scenarios, the Pluto system formed substantially closer to the Sun than its current heliocentric distance. The first modern explanation of its orbital configuration was by being swept up in resonance by an outward-migrating Neptune (Malhotra 1993, Malhotra 1995). This ``bottom up'' capture would have collected a low-excitation Pluto in the 3:2 resonance and carried it outward while exciting its orbit in the process. Later work showed that in order to achieve both the current inclination and eccentricity of Pluto's orbit, multiple epochs of resonant interactions would be required, including an early epoch of inclination excitation in the $\nu_{18}$ secular resonance (where the precession rate of a TNO's longitude of ascending node matches the precession rate of Neptune's longitude of ascending node; see, e.g., Morbidelli et al. 1995), followed by capture, transport, and eccentricity excitation by the 3:2 MMR, followed by further inclination excitation in the Kozai resonance (Malhotra 1998). This sequence of events would require Neptune to migrate by 5-10 AU in total (Hahn \& Malhotra 1999), and with Pluto capturing into resonance outside of $\sim28$ AU. Considering the full current resonant properties of Pluto, Gomes (2000) argued that these processes would generate Pluto-like objects given origin locations near 30.5 AU. However, the subsequent discovery of a very broad inclination distribution of more distant non-resonant ``classical'' KBOs presented a challenge: these inclinations could not have been excited by resonance sweeping alone, and required another explanation. A solution was proposed by Gomes (2003), where objects are initially excited and launched outward to current KBO distances by scattering encounters with Neptune, fueling its migration. These objects can experience transient phases of lower eccentricity due to resonant interactions, raising their perihelia away from Neptune, and then fall out of resonance as Neptune continues to migrate, effectively stranding them in stable high-inclination, relatively high-eccentricity orbits. In this scenario, resonances are populated in a ``top down'' way, where objects are generated with high inclination and eccentricity through scattering events and are later permanently captured into 3:2 resonance. This removes the need for the most extreme extents of smooth migration by Neptune, as the inclination and eccentricity of Pluto and its resonant neighbors are not due solely to excitation through resonant sweeping. This scenario was further explored and refined in Levison \& Morbidelli (2003) and Levison et al. (2008), which showed that even low-excitation orbits can be established by a history of scattering and transient resonant diffusion. These families of models (e.g., Gomes 2003, Levison et al. 2008) argued that Pluto and the rest of the TNO population were all moved outward after forming in a massive (10-50 M$_{\oplus}$) disk of planetesimals that formed substantially closer to the Sun, truncated somewhere between 30 and 35 AU. 

However, more recent work has demonstrated that Pluto does not share a common origin location with all extant TNOs. The low-inclination CCKBOs were shown to be very numerous, and to have a unique size-frequency distribution (Brown 2001, Levison \& Stern 2001, Bernstein et al. 2004, Fraser et al. 2010, Petit et al. 2011, Fraser et al. 2014), a unique color distribution (Trujillo \& Brown 2002) and albedo distribution (Brucker et al. 2009), and a unique population of widely-separated binary systems (Noll et al. 2008). All of these properties made the cold classical KBOs physically distinct from other TNO populations. Further, the wide binary systems are very sensitive to disruption by any epoch of Neptune scattering that would have implanted them in their current orbits (Parker \& Kavelaars 2010) and to collisional evolution as may have been expected during an early epoch inside a massive disk (Nesvorn\'{y} et al. 2011, Parker \& Kavelaars 2012). These factors together point to the CCKBOs having an origin distinct from Pluto and the rest of the TNO populations, likely forming in-situ in a low surface density extension of the primordial planetesimal disk and suffering very little orbital excitation or collisional comminution over its history (e.g., Cuzzi et al. 2010, Batygin et al. 2011). Indeed, there is an emerging consensus that the physical and dynamical properties of CCKBOs are consistent with having emerged directly and very rapidly from aerodynamically-enhanced collapse of pebble swarms (e.g., Nesvorn\'{y} et al. 2010, 2019) with little subsequent modification. 

Other than the CCKBO sub-population and a proportion of the resonant populations that were ``swept up'' from it, Pluto likely shares a common origin with most other identified TNO sub-populations. A caveat here is the relatively recently-identified ``distant detached'' population exemplified by Sedna. The origin of these objects remains contested, and while a number of scenarios exist where they emerged from the same planetesimal disk as Pluto (Brown et al. 2004, Morbidelli \& Levison 2004), others exist where they have unique source populations (Jilkova et al. 2015). Regardless, they are likely to have experienced substantially different orbital evolution over their lifetimes and their properties (including their propensity to host satellite systems; Sedna is the largest-known TNO with no known satellites) may differ from the other TNO populations as a result.

The most up-to-date modeling efforts indicate that Neptune migrated relatively slowly into a $\sim20 M_{\oplus}$ disk that extended out to 30 AU and contained several thousand (or $2-8 M_{\oplus}$-worth) Pluto-sized planetesimals (Stern 1991, Nesvorn\'{y} \& Vokrouhlicky 2016). This large initial population of Pluto-sized objects is required to introduce a ``grainy,'' discontinuous aspect to Neptune's migration to curtail the efficiency of direct resonant capture and increase the efficiency of resonant drop-off in the hot classical population, which otherwise would result in resonant populations far more numerous than observed and hot classical populations more anemic than obervered (Petit et al. 2011, Gladman et al. 2012, Nesvorn\'{y} 2015). While the initial population of proto-Plutos was very large, relatively few of them survived transport into the extant TNO populations. 

\begin{figure}[t]
    \centering
    \includegraphics[width=9cm]{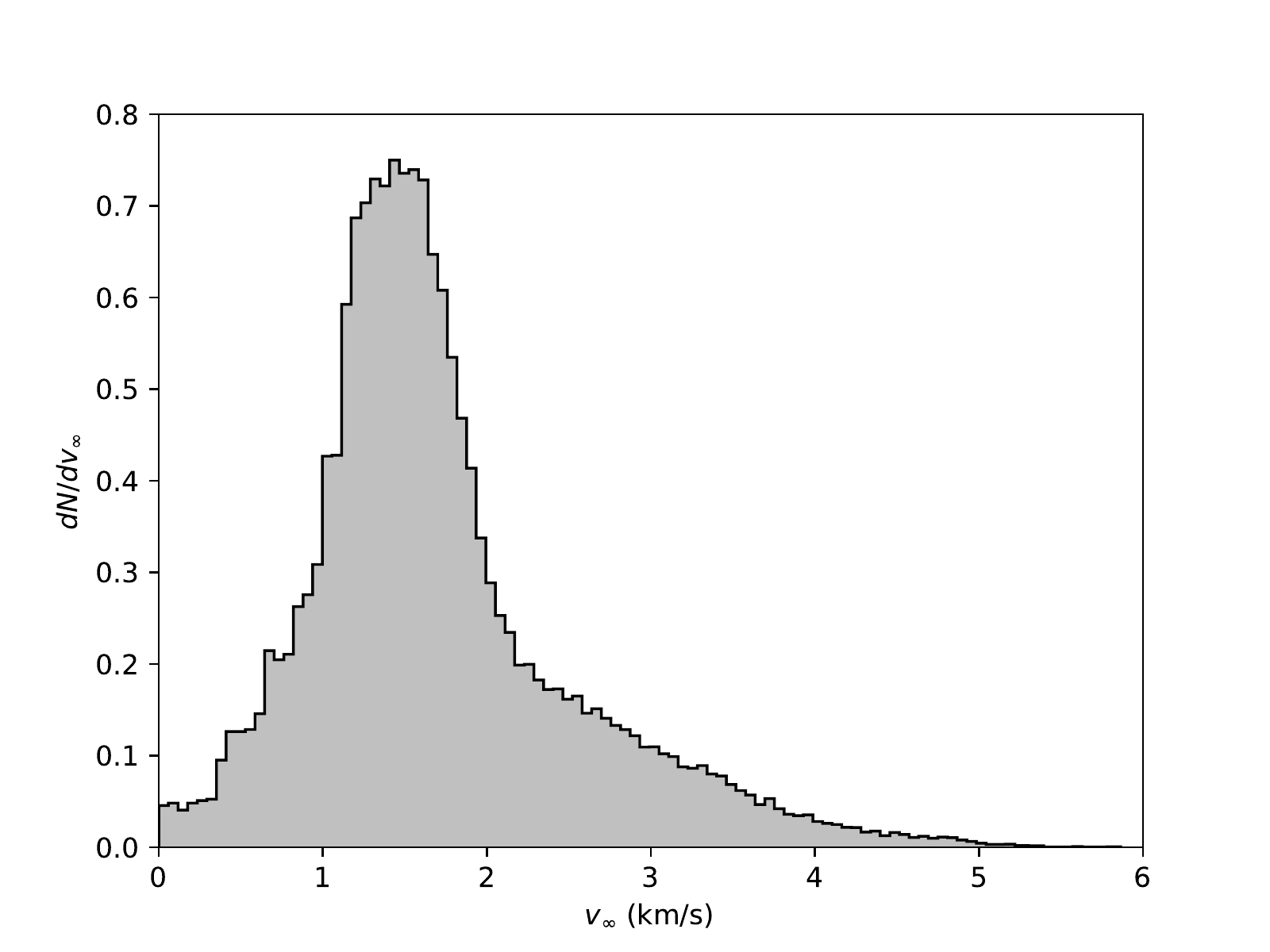}
    \caption{ Adopted $v_{\infty}$ distribution for impactors into the Pluto system, using a population-weighted average over all populations considered in Greenstreet et al. (2015). }
    \label{fig:speed}
\end{figure}

As Pluto's osculating orbital elements evolve cyclically due to the resonances it occupies, its intrinsic collisional coupling to each TNO sub-population changes. Long-term average collision rates were compiled by Greenstreet et al. (2015) for each of the most-populous well-characterized TNO sub-populations, including the hot and cold classical KBOs, several resonant KBO populations, and the scattered disk. For the population of potential impactors larger than $d\simeq 100$ km (the approximate location of the size-frequency distribtion's break to a shallower slope), the classical belt contributes a majority of all the impactors to the Pluto system (74.7\%), while the 3:2 resonant population comprises the next largest supplier of impactors (18.6\%). The typical impact speeds for objects in each population varies, and the amalgam velocity distribution weighted by the impact odds and population sizes is illustrated in Figure \ref{fig:speed}.

\section{\textbf{BUGS ON THE WINDSHIELD}}

One of the most direct ways that the Pluto system can provide insight into the properties of trans-Neptunian populations is by treating the surfaces in the system as witness plates. Most of these surfaces represent a record of the integrated history of bombardment of the Pluto system by the TNO sub-populations that it physically passes through. The order-zero property that can be extracted from this record is the sheer number of impacts recorded down to the limit of any given set of observations, while a first-order property is how the sizes of those craters are distributed. Higher-order properties, such as spatial correlation with geological units, two-point autocorrelation functions, and others, can tease out relative and absolute ages of surface units and whether or not a subset of the impactors bore satellites.  

\bigskip
\noindent
\textbf{2.1 Inferring the Size-Frequency Distribution of small TNOs}
\bigskip

The size-frequency distribution (SFD) of TNOs is a powerful tracer of the fundamental processes of planetesimal assembly before and during the era of planet formation, as well of any subsequent collisional evolution over the age of the solar system. A variety of means have been used to estimate the SFD properties of TNO sub-populations, including so-called ``direct'' observation in reflected disk-integrated light, occultations of background stars, and inferring the impactor populations responsible for generating craters on the surfaces of outer solar system worlds. It is worth briefly considering the physical properties that each of these methods are actually sensitive too, and the strengths and shortcomings of each.

First, it is useful to identify those properties of TNOs that are most likely to be directly influenced by their formation processes versus those that are merely accessory to them. Most fundamentally, growth processes partition \emph{mass} between forming bodies. Disk-integrated luminosity of a TNO is the result of a combination of factors including its mass, density, shape, orientation, and surface albedo. Occultation cross section is the result of mass, density, shape and orientation. In the point-mass approximation and in the gravity regime, craters typically scale with mass of the impactor, velocity of the impactor, density of the impactor, and density of the target. 

With no substantial sensitivity to albedo, shape, or orientation distributions, the size-frequency distribution of impact craters provides a powerful and complementary measure of this most fundamental mass distribution. However, the inferences that can be drawn from impact crater populations are confounded by a distinct set of factors, including uncertainties in crater scaling laws, the properties of the target bodies, the generation of secondary craters, modification of crater populations by geologic processes, and limitations in the calibratability of current-generation crater surveys. While these factors must be accounted for, assays of crater populations remain one of the most powerful probes of the fundamental partitioning of mass between small bodies in trans-Neptunian space, and will likely remain so until serendipitous stellar occultation surveys increase their yields by several orders of magnitude. In the end, a \emph{consensus} understanding of the TNO populations and their evolution through time should be reached across all methods for studying them. If, for example, stellar occultations surveys and impact crater surveys produce persistently discrepant results, we are potentially missing something fundamental and the tension between methods may be highlighting the path to a deeper understanding of the fundamental properties of outer solar system populations. 

Analysis of the size-frequency distribution of craters in the Pluto system suggests that small TNOs are less numerous than would be anticipated if they had ever reached collisional equilibrium (Singer et al. 2019). If the size-frequency distribution of impactors is a scale-free power law $dN/dD_{i} \propto D_{i}^{-q_{i}}$, then the size-frequency distribution of simple craters in the gravity regime will also be a scale-free power law over a large range of sizes, $dN/dD_{c} \propto D_{c}^{-q_{c}}$. Given the Housen, Holsapple, and Schmidt scaling factor $\mu$ (e.g., Holsapple 1993), we can determine the slope of the impactor distribution implied by the crater distribution:

\begin{equation}
q_{i} = \left(1 - \frac{\mu}{2 + \mu}\right) q_c + \frac{\mu}{2 + \mu}
\end{equation}

For typical experimentally-derived values of $\mu$ between 0.41 and 0.55 (Holsapple 1993), this implies an impactor size-frequency distribution slightly \emph{shallower} than the measured crater size-frequency distribution. The physical extremes for the $\mu$ parameter are $\frac{1}{3}$ and $\frac{2}{3}$, and correspond to pure energy conservation and pure momentum conservation. Both extremes result in impactor SFD slopes shallower than crater SFD slopes. Thus, absent any completeness effects or other non-idealities, the $q_c \sim 1.8$ slope for craters smaller than 10 km on Charon's Vulcan Planum measured by Singer et al. (2019) implies an impactor slope $1.6 \lesssim q_i\lesssim 1.7$ at the extreme limits of $\mu$ for impactors in the size range $0.1 \mbox{ km} \lesssim d_i \lesssim 1 \mbox{ km}$.

However, the Pluto system's crater size frequency distribution suggests that over the range of crater diameters recorded, the impactor size-frequency distribution is \emph{not} a scale-free power law. There is an apparent break in the crater size frequency distribution at a diameter of $d_c \sim 10-15$ km. To fully describe the state of knowledge of the properties of the small-TNO size-frequency distribution as informed by the crater population on Pluto and Charon, and to further fold in existing knowledge of the size frequency distribution at larger sizes from direct-detection surveys, a more sophisticated analysis is required that permits marginalization over nuisance parameters like $\mu$ and any parameters that encode sensitivity and completeness of the crater census. The following outlines such an approach and applies it to existing crater catalogs to estimate the current state of knowledge regarding the size-frequency distribution of small TNOs.

\bigskip
\textit{2.1.1 Likelihood-Free Inference}
\bigskip

Cratering is a complex process, and to use a measured distribution of crater properties from one or several surface units to infer the intrinsic properties about the population of projectiles that generated them requires a great deal of statistical care. The complexities of the cratering process and the physical task of conducting a census of a crater population are difficult to account for in standard likelihood-based inference. However, relatively recently there has been a surge of interest in and development of \emph{likelihood-free} Bayesian inference. While many of these methods have their origins in the genomics community, their application in astrophysics (Ishida et al. 2015, Hahn et al. 2017, Hsu et al. 2018, Witzel et al. 2018, Sandford et al. 2019) and planetary science (Parker 2015, Mazrouei et al. 2019) has been growing in recent years. 

The chief advantage of likelihood-free inference for planetary science is that it enables existing forward-modeling infrastructure to be applied within a Bayesian framework. More concretely, likelihood-free inference enables rigorous Bayesian inference using stochastic generative models, and it can be far more straightforward to encode all the complexities of planetary processes and observations within a stochastic model than it is to write down the formal likelihood integral that captures the interplay of these processes, let alone actually \emph{evaluate} that integral.

The following section outlines a likelihood-free framework for Bayesian inference of the parameters of several models of the population of small TNO projectiles that have been impacting the Pluto system over the age of the solar system. This framework and approach is generic and can be applied to other planetary surfaces or to other planetary science population synthesis challenges. 

The primary likelihood-free inference tool used here is an Approximate Bayesian Computation rejection sampler (ABCr; Pritchard et al. 1999, see Marin et al. 2011 for methods review), one of the simplest likelihood-free inference methods to implement and understand. ABCr replaces likelihoods with a distance metric $D_{ABCr}$ measured on a set of summary statistics extracted from both the real observed data sample and a synthetic ``observed'' sample. In a single ABCr trial, a set of parameters describing the model are drawn from their prior distributions, and a single synthetic ``observed'' sample is drawn from the model with those selected parameters. The distance metric is computed for that trial and recorded along with the parameters that defined that instance of the model. After many trials, the sets of parameters that happened to produce the smallest distances between synthetic and observed data are kept, and the rest rejected (typically, a 0.1\%-percentile threshold is used). With a carefully chosen distance metric, a small enough threshold for retention, and a large enough number of trials, the retained set of model parameters will approximate their posterior distribution given the observed data. This work adopts a distance metric that is the sum of Kolmogorov-Smirnov $D_{KS}$ statistics over the observables in question. This is similar to the treatment in Parker (2015) for Neptune Trojan orbit distribution parameter estimation and Mazrouei et al. (2019) for estimating the parameters of a time-varying impact rate model for the Earth and Moon.

Further, ABCr can be used to compute a Bayes factor for one model out of a suite of models. It proceeds as before, but in any given trial, first a model is selected at a rate commensurate with its prior probability, then the parameters for that model are proposed. After completing many trials and retaining the best 0.1\%, ratio of the number of retained trials of each model to the number of retained trials of other models in the final sample is the Bayes factor in favor of that model. That is, if 95 instances out of 100 retained trials were Model A, and 5 were Model B, the Bayes factor is 19:1 in favor of Model A. 

\bigskip
\textit{2.1.2 Measuring the Population of Craters}
\bigskip

Unlike direct-detection surveys of the TNO population, the surfaces on which crater censuses are conducted are often not available for re-visiting with new or different observational strategies once an initial census is completed. The Pluto system has been visited by a single spacecraft, which is likely to be the only time the system is visited for at minimum a decade. The observational datasets collected during flyby -- not informed by any information about any evidence for novel properties of the TNO population that may be encoded in the system's surface units -- are the only dataset from which we are able infer any properties about the crater populations on Pluto and its moons. It is of particular importance to avoid overfitting the data at our disposal because it will be impossible to test with an independent dataset for a very long time.

Further, craters are imprinted on surfaces with substantial albedo and topographic variation, and are themselves subject to endogenic modification processes that may vary from unit to unit on a given surface as well as from world to world. Calibrating the detection efficiency of any crater census is vastly more difficult than calibrating a direct detection survey, which can generally be thought of as a collection of stationary and moving point sources on a relatively smooth and featureless background that varies in a predictable way, with overall sensitivities governed by photon and detector noise. Algorithmic and human limitations can be tested for by injecting simulated targets into the same dataset under consideration and estimating the fraction of those synthetic targets recovered by the analysis as a function of any parameter of interest (speed of motion, luminosity, color, and so forth). Even so, pushing into new regimes of luminosity has resulted in some substantial mischaracterizations of the population of small TNOs in the past (e.g., Cochran et al. 1995 \& 1998, Brown et al. 1997, Gladman et al. 1998) due to unaccounted-for noise sources in the datasets at hand. These noise sources are generally identified in subsequent analysis of the original dataset, and new datasets are gathered to confirm these analyses.

In some cases, there have been large direct-detection datasets for which no complete discovery bias calibration was performed at the time of discovery. Parker (2015) developed a \emph{survey-agnostic} method of calibrating a set of these datasets post facto, where parameters of a model of discovery biases were marginalized over to uncover what information was still available in the data themselves about the population under consideration, \emph{given} our uncertainty in these discovery biases. With some adjustment, these methods can be applied to crater populations to better assess our true state of knowledge about the population of projectiles that generated them. 

\begin{figure*}[t]
    \centering
    \includegraphics[width=15cm]{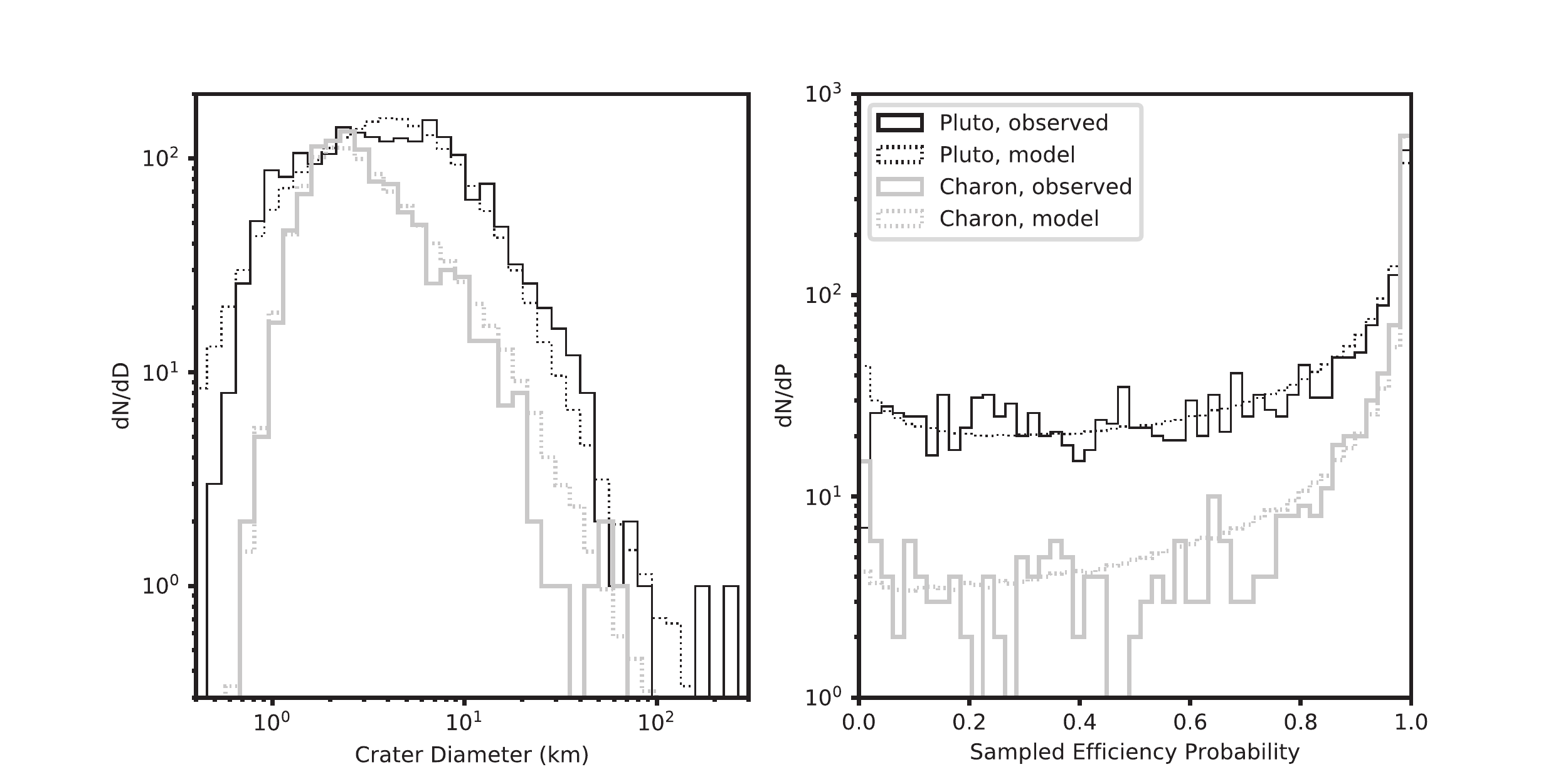}
    \caption{ Differential histograms of crater diameters (left) and sampled efficiency function (Eqn. \ref{eqn:eff}) probabilities (right) for observed (solid) and model (dotted) samples from Pluto (black) and Charon (gray). Illustrated model parameters are $a_p=0.7$, $b_p=-4.5$, $a_c=0.6$, $b_c=-11$, $q_1=8.5$, $q_2=2.9$, $q_3=1.85$, and $D_b=10$ km.}
    \label{fig:SFD_obs}
\end{figure*}

The majority of crater censuses are conducted via analysis by the human eye. As with any human-in-the-loop process, there is a degree of subjectivity informed by experience with past crater counting efforts present in any crater identification. Ideally, this subjectivity would be controlled for through calibration, perhaps by generating synthetic datasets with the same topographic and albedo variations as the target under consideration, injecting simulated crater populations into them, and determining the properties of the recovered sample as a function of both the injected population properties and the human researcher conducting the analysis. At present, this level of calibration is largely impractical. While such \emph{absolute} calibration may be absent, the variation \emph{between} human researchers has been tested (Robbins et al. 2014) for a \emph{fixed set of crater population parameters and terrain properties}. Thus, it is possible to estimate how different the estimated parameters for a given crater population might be as determined by any two human researchers, but \emph{not} always possible to estimate how much they both err from the underlying truth.

There are two crater catalogs available for the Pluto system, both drawn from New Horizons imaging datasets; the amalgam catalogs from Robbins et al. (2017, hereafter R17) that combines all available datasets over the entire imaged areas of Pluto and Charon, and the unit-by-unit, image-by-image catalogs of Singer et al. (2019, hereafter S19). The two datasets are not fully independent as they are drawn from the same data and have contributions from common researchers, but they do have properties unique from one another that makes them each valuable for separate analyses. The R17 catalogs cover the entire imaged portions of both Pluto and Charon, while S19 focuses on a smaller subset of Pluto and Charon for which high-resolution information is available; R17 provides a unique identifier each crater, while S19 does not indicate duplicate craters if they appear in multiple imaging datasets; R17 amalgamates inputs from multiple human researchers and includes an estimate of the average subjective confidence level reported by those researchers without further thresholding, while S19 represents the best-effort dataset produced by a single researcher that pass a fixed threshold confidence level; and S19 reports crater populations resolved into different image subsets and terrain units, while R17 reports a single disambiguiated population of craters per body without further distinction based upon terrain or image source. As an additional refinement step, S19 used topographic data to confirm features where such topography was available. The following analyses use the R17 dataset for parameter estimation and the S19 dataset for validation. 

To ensure a high-quality sample of craters from well-observed regions on both Pluto and Charon, only those craters from the R17 catalog with a reported confidence level of three or greater (one a one-to-five subjective scale, with three corresponding to a better than 50-50 odds that a feature is in fact a crater; see R17 for further details regarding this catalog) in the original catalog are selected, and from that sample only those which are in the upper 50th-percentile of regional crater density (as defined by the distance to the 100th nearest neighbor of each crater) are considered. This rejects craters from the encounter farside of both bodies and from sparsely-cratered regions; because their neighbors cover large fractions of each body, estimating the local detection efficiency from the nearest neighbors as described in the following section would be innacurate and thus they are removed from the sample. This results in a winnowed sample of 1,011 craters on Charon and 2,016 craters on Pluto for SFD parameter estimation. The final sample extracted from R17 for subsequent analysis in this work is illustrated in Figure \ref{fig:SFD_obs}.

\begin{figure*}[t]
    \centering
    \includegraphics[width=12cm]{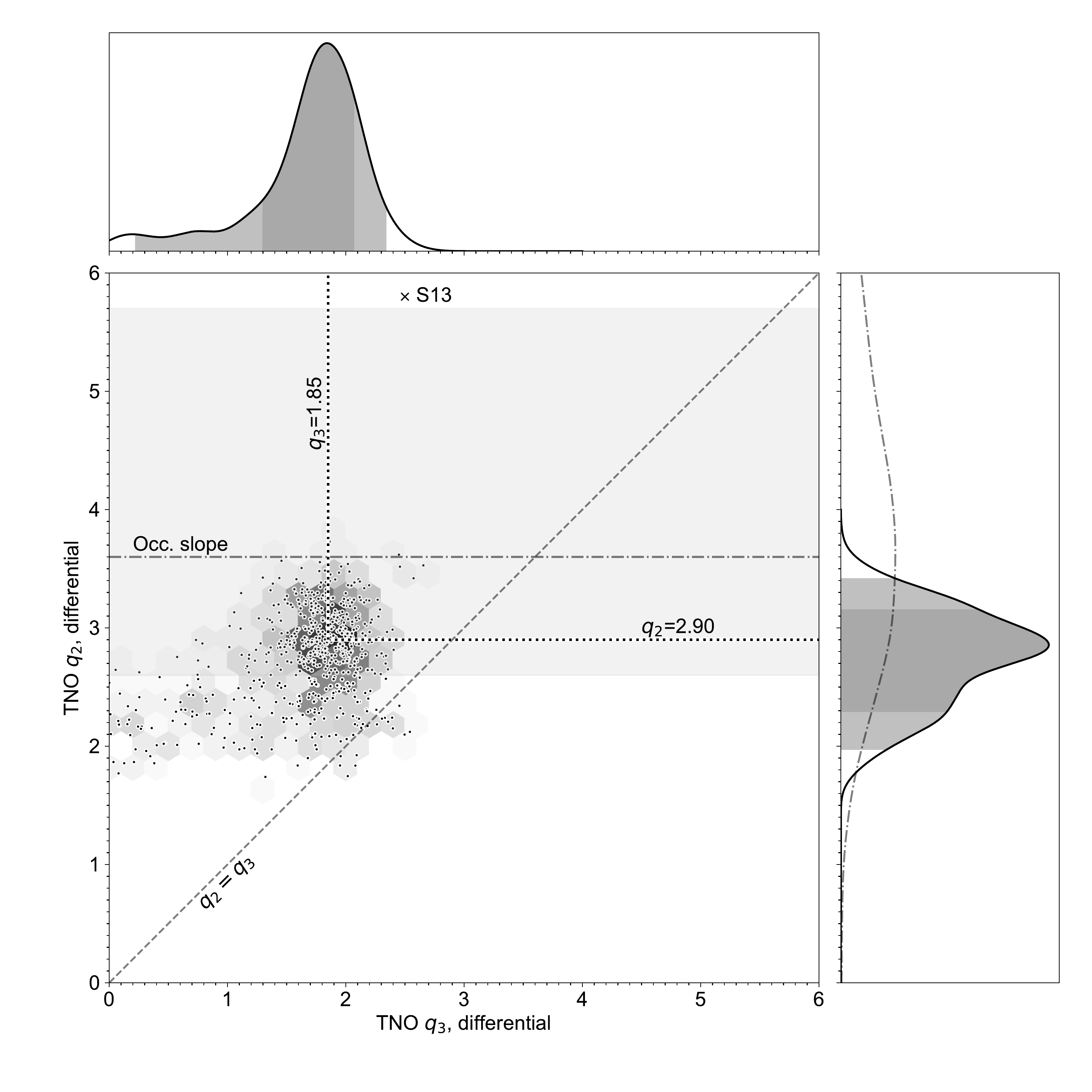}
    \caption{ Marginalized posterior PDFs for TNO size-frequency distribution power-law slopes; $q_3$ applies to the smallest objects that created craters on Pluto, $q_2$ applies to the largest objects that created craters on Pluto and is the small-object slope for directly-detected KBOs. $q_1$ (not shown) is measured for the largest TNOs by direct detection, and is population dependent (see Fraser et al. 2014). Proposed differential slopes for the relevant size regimes from Schlichting et al. (2013; S13) shown for comparison. Gray bar and dot-dash line show the PDF of mid-size slope inferred from occultation detections alone.}
    \label{fig:SFD1}
\end{figure*}

\bigskip
\textit{2.1.3 Modeling Crater Creation and Detection}
\bigskip

Our understanding of the processes by which a projectile impacting a planetary surface generates a crater of given properties has a long history, and is informed both by laboratory experiments and observational inference. A full summary of this history is beyond the scope of this paper; for a review, see Holsapple (1993) and for recent considerations regarding impacts into high-porosity surfaces see Housen et al. (2018). Briefly, during an impact, the kinetic energy of a projectile is partitioned into heat -- which can melt or vaporize target and projectile material -- and into the mechanical work of compressing, excavating, and ultimately ejecting material from target surface. A huge range of variable factors influence the properties of the final crater generated by an impact, including the size, speed, and impact angle of the projectile, the bulk material properties of the projectile and of the target surface, and the surface gravity of the target body. The results of general impactor-to-crater scaling laws generally produces a relationship with the final diameter of the crater being proportional to the diameter of the impactor to a power near unity. The constant of proportionality varies depending on all the properties described above, and -- importantly -- the exponent depends on several of these properties as well. 

It is typical to work backwards in impactor population synthesis, where the properties of a given crater or population of craters are mapped back to the properties of an impactor population. However, to accomplish this, a great deal of information must be discarded. Any given crater could have been produced by impactors of a variety of sizes, impact speeds, and impact angles -- but a given impactor, with its fixed size, speed, and trajectory, can only produce one particular crater. That is to say, impactors make craters, but craters do not make impactors. \emph{Forward}-modeling this process can therefore retain the influence of the distribution of impactor properties \emph{other} than size all the way to the final marginalization step, instead of averaging over them in the crater scaling relation. For simplicity of the current analysis, the cratering model of Zahnle et al. (2003) is adopted as a baseline. That is,

\begin{equation}
    D_{s} = 13.4 (v_i^{2}/g)^{0.217} (\cos(\theta) \rho_i/\rho_t)^\frac{1}{3} d_i^{0.783} \mbox{ km},
\end{equation}

for $D_s \leq D_c$, where $D_s$ is the rim-to-rim diameter of a simple crater smaller than the transition to complex craters $D_c$; to account for craters larger than $D_c$, the final crater diameter $D_f$ is given by

\begin{equation}
D_{f} = 
 \begin{cases} 
      D_s & D_s \leq D_c \\
      D_{s} (D_s / D_c)^{0.108} & D_s > D_c 
 \end{cases}
 \label{eqn:final_crater}
\end{equation}

The exponent in the second case is slightly lower than that used in Zahnle et al. (2003) and is based on McKinnon and Schenk (1995). Here $v_i$ is the impact speed in km/s, $g$ is the target body's escape speed in cm/s$^2$, $\rho_i$ and $\rho_t$ are the bulk densities of the impactor and target materials (respectively) in grams/cm$^3$, $d_i$ is the impactor diameter in km, and $\theta$ is the impact angle from normal. The exponent on $d_i$ implies a $\mu$ parameter of 0.55, and this scaling law is appropriate for generating crater diameters in the gravity-dominant regime for non-porous target materials. $D_c=10$ km as adopted for both Pluto and Charon, based on crater morphology transition (Robbins 2019, pers. comm.).

The intrinsic size-frequency distribution of the impactor population is taken to be a piecewise power-law distribution;
\begin{small}
\begin{equation}
    p(d) \propto 
 \begin{cases} 
     \left(\frac{d}{d_{b,0}}\right)^{-q_{0}} & d_{b,0} \leq d < d_{b,1} \\
     \left(\frac{d}{d_{b,0}}\right)^{-q_{i}} \times \prod_{j=1}^{i} \left(\frac{d_{b,j}}{d_{b,0}}\right)^{q_{j} - q_{j-1}}  & d_{b,i} \leq d < d_{b,i+1}
 \end{cases}
\end{equation}
\label{eqn:sfd}
\end{small}

where $d_{b,0}$ is the smallest size to be considered (must be $>0$), the the rest of the list $d_{b,1}...d_{b,n-1}$ are the diameters of $n-2$ proposed breaks in the distribution, and $d_n$ is the largest size to consider (or $+\infty$).

To model a synthetic ``observed'' crater population, first a sample of impactor diameters $d_i$ is drawn from a proposed size-frequency distribution using Eqn. \ref{eqn:sfd}. Then a sample of impact speeds at infinity $v_{\inf}$ is drawn from the Greenstreet et al. (2015) distribution for the Pluto system (Figure \ref{fig:speed}), and augment it with the escape speed of the target body to determine the impact speed of each impactor $v_i = \sqrt{v_{\inf}^2 + v_{esc}^2}$. Finally, a sample of impact angles $\theta$ is drawn from a distribution uniform in $\sin^2(\theta)$ over $0^\circ \leq \theta \leq 90^\circ$. Running these samples through Eqn. \ref{eqn:final_crater} results in a proposed sample of craters on the surface of the target, but do not yet include observational sampling effects. 

To model observational effects, a functional form for observational completeness $\eta$ is adopted which is inspired by the rollover functions used in direct-detection surveys. For crater $i$ with observed diameter $D_{o,i}$,

\begin{equation}
    \eta(D_{o,i} | D_{m,i}) =   \frac{1}{2} \left(1 - \tanh\left[ b \log_{10} \left( \frac{D_{o,i}}{a D_{m,i}} \right) \right] \right)
\label{eqn:eff}
\end{equation}

\quad

\noindent where $a$ and $b$ are (nuisance\footnote{A nuisance parameter is a model parameter that is not of direct physical interest but which has some potential to impact the probability density function (PDF) of parameters that are of physical interest}) scaling parameters that apply to the entire population of a target object and which are marginalized over, and $D_{m,i}$ is the median size of the 100 craters nearest to the crater under consideration. This sample size is somewhat arbitrary, and 100 craters was chosen to balance regional fidelity with good signal-to-noise properties based on the sample sizes in R17. Functionally, this model of observational completeness assumes that completeness is regional on Pluto and Charon, and that the observed SFD of craters in a given region encodes both the underlying intrinsic SFD and a multiplicative efficiency function. Every crater on Pluto and Charon is assumed to represent a distinct sample from this selection process. The observable property \emph{recorded} for every crater that encodes this information is the median size $D_{m}$ of craters in its vicinity, \emph{not counting itself}. Thus, any given crater does not influence the estimation of the detection efficiency function that applies to it. The parameters $a$ and $b$ map the measured $D_m$ values to an efficiency function that applies to the diameter $D_o$ of the crater for which each $D_m$ was measured; $a$ sets the efficiency rollover width, and $b \times D_m$ sets the efficiency rollover diameter.  These two nuisance parameters (per target; four total over Pluto and Charon) completely describe the observational completeness effects as modeled, and their impact on the parameters of interest are marginalized over in the ABCr process.

When mapping a sample drawn from Eqns. \ref{eqn:sfd}--\ref{eqn:final_crater}, this produces a proposed sample of craters that may exist on Pluto. To map this to a proposed synthetic observed sample, one synthetic crater is selected from this sample for each measured $D_m$ value recorded from the observed sample, applying a weight to the proposed craters based on the implied efficiency function for each unique $D_m$ given a set of proposed $a$ and $b$ values. 

At this point, all the parts needed to conduct ABCr inference on the model parameters are ready. $10^7$ ABCr trials were conducted using a three-slope power law; the smallest-object slope $q_3$ was drawn from a uniform prior over [0, 3], the intermediate-size $q_2$ from the posterior PDF of ``cold'' population faint-object slopes from Fraser et al. (2014), and the large-object slope $q_1$ from the posterior PDF of ``cold'' population bright-object slopes from Fraser et al. (2014). The break between $q_3$ and $q_2$ was selected uniform in $log_{10}(D_b)$, over the range $0.1 \mbox{ km} \leq D_b \leq 100 \mbox{ km}$. The values of $a$ and $b$ were proposed separately for Pluto and Charon, each drawn from uniform priors over [0.1, 1.0] and [1, 100], respectively. In each trial, the two-sample KS-test statistics were retained for the proposed Pluto crater sample and the observed Pluto crater sample, the proposed Charon crater sample and the observed Charon crater sample, the proposed efficiency function values for every proposed Pluto crater and for every observed Pluto crater, and the proposed efficiency function values for every proposed Charon crater and for every observed Charon crater. These statistics were summed for each trial to generate the ABCr $D_{ABCr}$ metric. The latter two KS-tests permit the simultaneous assessment of whether the efficiency function model is well-matched to the data (that is, synthetic craters are not being selected from vastly different parts of the proposed efficiency function than observed craters were) and whether the size-frequency distribution is well-matched to the data -- see Figure \ref{fig:SFD_obs} for example distributions. After the $10^7$ trials were complete, the sample of proposed parameters that resulted in $D_{ABCr}$ values in the lowest 0.1\% of all sampled $D_{ABCr}$ values was extracted. This subsample is adopted as the posterior PDF for the parameters under consideration.

\bigskip
\textit{2.1.4 Size-Frequency Distribution Parameters}
\bigskip

The posterior PDF distributions for $q_2$ and $q_3$ are illustrated in Figure \ref{fig:SFD1}, marginalized over the efficiency function nuisance parameters ($a_{p}$, $b_{p}$, $a_{c}$, $b_{c}$) and the break location $D_b$. The mode value of the PDF and marginalized 68\% confidence intervals for these parameters are $q_{3} = 1.85^{+0.22}_{-0.55}$ and $q_{2} = 2.90^{+0.26}_{-0.61}$. These slopes are broadly consistent with those determined for the crater distribution in Charon's Vulcan Planum by Singer et al. (2019), but the addition here of the observational completeness modeling brings the SFDs expressed by both Pluto's and Charon's surfaces into agreement. 

While the data strongly prefer a three-slope model over a two-slope model, the location of the break between $q_2$ and $q_3$ is relatively poorly constrained to $D_b = 2.8^{+26}_{-2.2}$ km (mode and 68\% interval). This suggests that while a single slope for all TNOs smaller than $d \sim 100$ km is certainly a poor representation of the data, a sharply-broken two slope model may also be an inadequate description. If the population instead slowly transitions from one regime to another, then a single break diameter will not reproduce the data well. It is also possible that the assumption of common crater scaling relationship parameters for both Pluto and Charon is not supported by the data; if the exponent $\mu$ is slightly different for Pluto and Charon, indicating perhaps a difference in surface material properties, any residual tensions between the observed slopes on Pluto and Charon could be relaxed and possibly enable better constraint of $D_b$. Future work is merited that considers a broader family of SFD functional forms and crater scaling relationships.

\begin{figure*}[t]
    \centering
    \includegraphics[width=12cm]{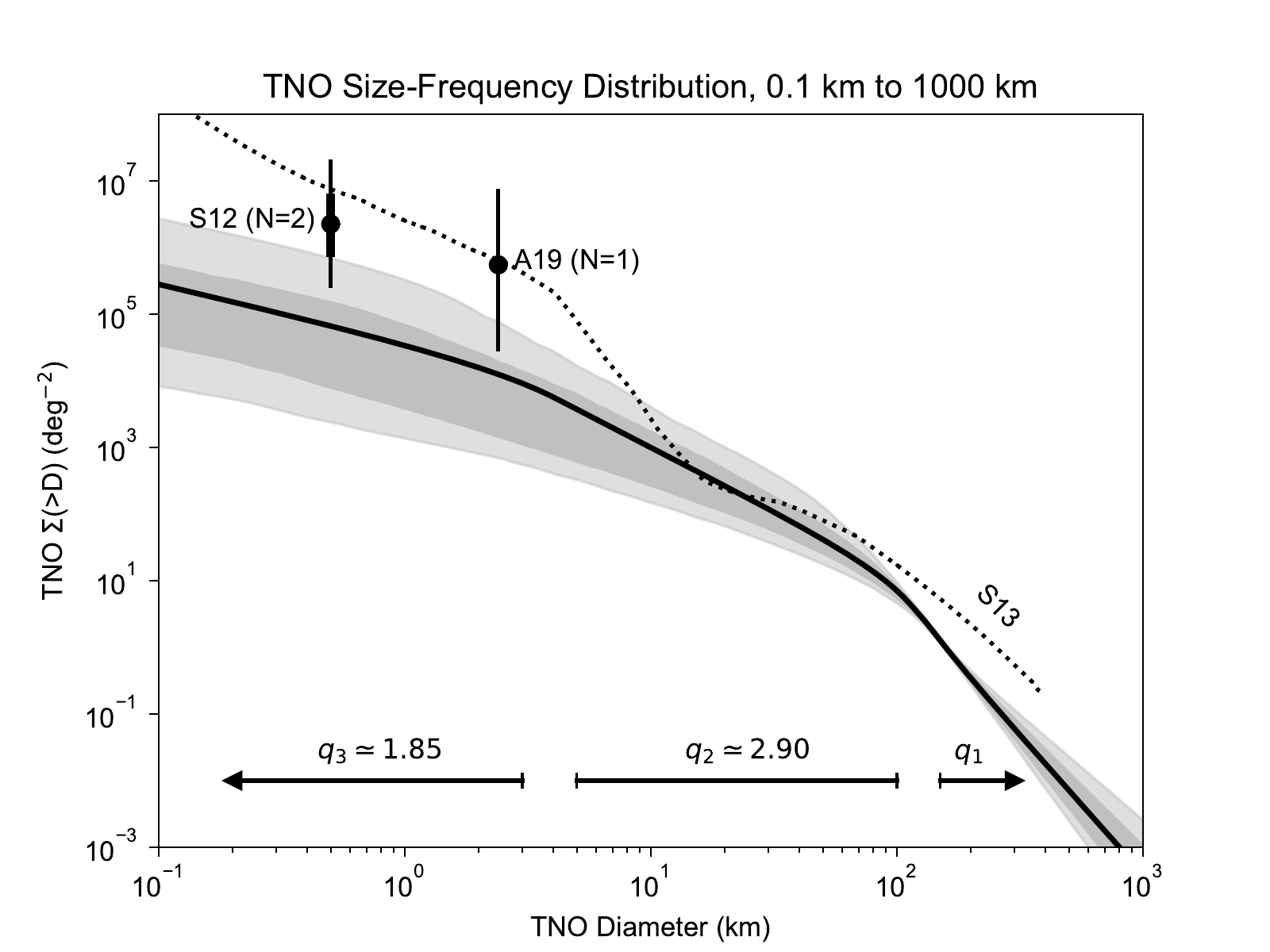}
    \caption{ Cumulative ecliptic surface density of TNOs, combining Fraser et al. (2014) luminosity function results and Pluto system crater population synthesis from this work. 68\% and 95\% confidence intervals illustrated by shaded regions. Surface density combines both models for ``hot'' and ``cold'' populations from Fraser et al. (2014). Nominal differential slopes for mid- and small-size regimes are illustrated; large object slope $q_1$ is population dependant. Surface densities inferred from occultation detections in Schlichting et al. (2012; S12) and Arimatsu et al. (2019a; A19) are also illustrated; S12 results have been revised down by a factor of two to account for an additional diameter-dependent term (see text). The theoretical SFD proposed in Schichting et al. (2013; S13) is illustrated with the dotted line for comparison; the inferred population does not support the sharp upturn at $d\sim10$ km.}
    \label{fig:SFD2}
\end{figure*}

\bigskip
\textit{2.1.5 Odds of the Gaps: Bridging Craters, Occultations, and Direct Detections}
\bigskip

Three serendipitous occultation events have been claimed that have moderate confidence (with estimated false-alarm probabilities of $p \lesssim 0.02-0.05$ per event) and were obtained at cadences better than half the Nyquist rate for TNO occultations (Nyquist $\sim 30$ Hz, Bickerton et al. 2009). Two were single-telescope event detections made with the Hubble Fine Guidance Sensor (Schlichting et al. 2009, Schlichting et al. 2013) and one was a dual-telescope event made with small ground-based telescopes (Arimatsu et al. 2019a). These events were observed at low to moderate ecliptic latitudes (+6.6$^\circ$, +14$^\circ$, and +8$^\circ$ respectively), and are inferred to represent candidate occultations of background stars by TNOs with diameters of 1.0 km, 1.1 km, and 2.6 km, respectively. 

A theoretical SFD was proposed by Schlichting et al. (2013) that would account for the inferred very large population of $d=1$ km TNOs in the Schlichting et al. (2009, 2012) occultation analyses. The relevant slopes are illustrated in Figure \ref{fig:SFD1} and the implied SFD shape is illustrated in Figure \ref{fig:SFD2}. Most notably, it is not the small-object slope $q_3$ that is most dissimilar from the population of TNOs inferred from the Pluto and Charon crater populations, but rather the \emph{mid-size} slope $q_2$, which corresponds to the very well-sampled regime covered by the largest craters on Pluto and Charon. It is also worth noting that the largest uncontested craters on Pluto and Charon were likely generated by impactors with sizes comparable to the current limit of direct-detection techniques (e.g., the diameter of 486958 Arrokoth. Thus, a very steep slope in this intermediate size regime should be immediately obvious in the Pluto and Charon crater data, at sizes for which craters are most difficult to miss observationally and most difficult to erase geologically -- from Figure \ref{fig:SFD2} it appears that 90--99\% of craters generated by $0.1 \mbox{ km} \leq d \leq 4 \mbox{ km}$ would need to have been erased in a size-independent way. This seems deeply unlikely, and the Schlichting et al. (2013) model is not supported by the crater populations on Pluto and Charon (Singer et al. 2019). 

However, how discrepant are the actual inferred populations drawn directly from observations? The shadow diameter of a Fresnel regime shadow of a spherical object (defined across the width of the first Airy ring;  Nihei et al. 2007) is $W(d) \simeq \left( (2 \sqrt{3} F)^{3/2} + d^{3/2} \right)^{2/3}$, where $F=1.3 km$ is the Fresnel scale for an occulter at 40 AU in optical wavelengths. The size dependence of this shadow is not accounted for in the population inference calculations of Schlichting et al. (2012), resulting in an overestimate of the population by a factor of $\sim2$. The revised quantity is illustrated in Figure \ref{fig:SFD2} along with the estimates from the Arimatsu et al. (2019a) occultation. While these inferred TNO populations are larger than those indicated by the crater populations, the large uncertainties in both the crater-inferred population and the much smaller occultation datasets do not immediately rule one another out.

Further, the relative probability of detecting an occultation as a function of occulter diameter $d$ in a given survey and at a fixed heliocentric distance is given by

\begin{equation}
    P(d|q) \propto \eta(d)W(d)d^{-q},
\end{equation}

where $q$ is the local size-frequency distribution slope and $\eta(d)$ is the detection efficiency of a given survey for an object of diameter $d$. Given the efficiency function published in Schlichting et al. (2012), there is a large discovery volume at sizes smaller than the two claimed detections. The peak detection probability should peak at sizes much smaller than the claimed detections if the local size-frequency distribution slope is steep. The Arimatsu et al. (2019a) detection is much closer to their claimed detection limit, consistent with the expectations of a somewhat steeper slope in the size range covered by that survey ($d\gtrsim2$ km). Taking these three candidate detections, their respective confidence levels, and the two surveys' detection efficiency functions, an average slope of $q \sim 3.6^{+2.1}_{-1.0}$ (mode and 68\% interval) is most consistent with the data across the size range covered by these surveys ($0.4 \mbox{ km} \lesssim d \lesssim 2$ km). This slope, derived from the occultation surveys themselves and requiring no additional step of inference to connect them to the full population of TNOs, prefers values slightly steeper than the $q_2$ mid-size slope inferred from the population of craters on Pluto and Charon, but the two values are not formally inconsistent.

This suggests that, if anything, something might be somewhat amiss with the \emph{normalization factors} that connect the three measurement techniques -- direct detection surveys, occultation surveys, and inference from crater populations. Internally, the local slopes are in reasonable agreement, but the sizes of populations inferred do not seem to match yet. With improved orbit distributions from OSSOS (Bannister et al. 2016), improved stellar diameters from \emph{Gaia} (Stevens et al. 2017), improved understanding of the influence of non-spherical shapes on occultation profiles (Castro-Chac\'{o}n et al. 2019), and improved direct-detection size-frequency distributions for the smallest TNOs observed in reflected light (Parker et al. 2017), it may be the case that the populations inferred from these three methods can be brought comfortably into agreement. If not, more exotic explanations may be required, such as drastic changes in material density, extremely unusual population-specific size-frequency distributions, or as-yet unimagined geologic processes on Pluto and Charon.

\bigskip
\noindent
\textbf{2.2 Doublet Craters and the TNO Binary Population}
\bigskip

\begin{figure*}[t]
    \centering
    \includegraphics[width=8cm]{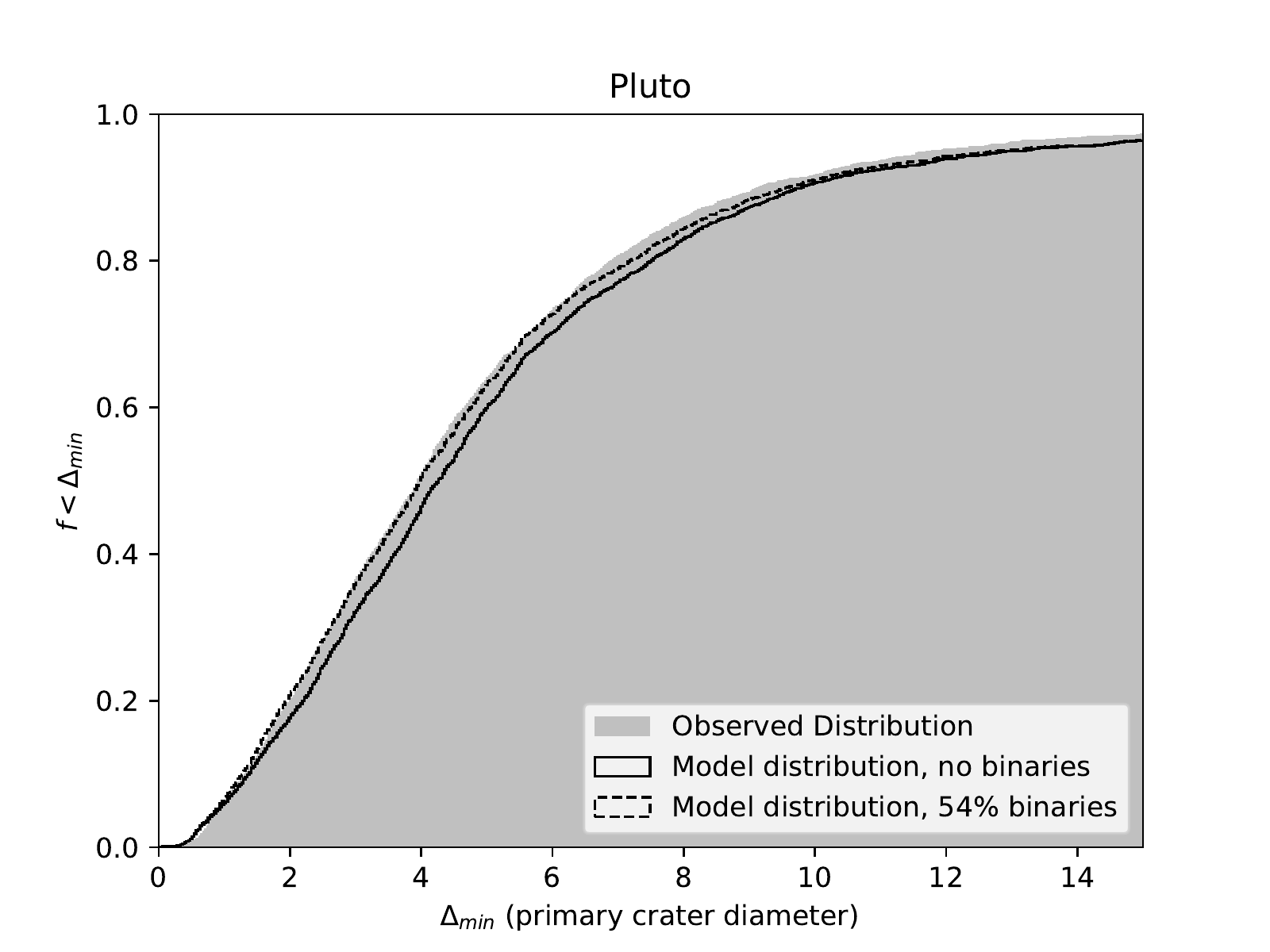}
    \includegraphics[width=8cm]{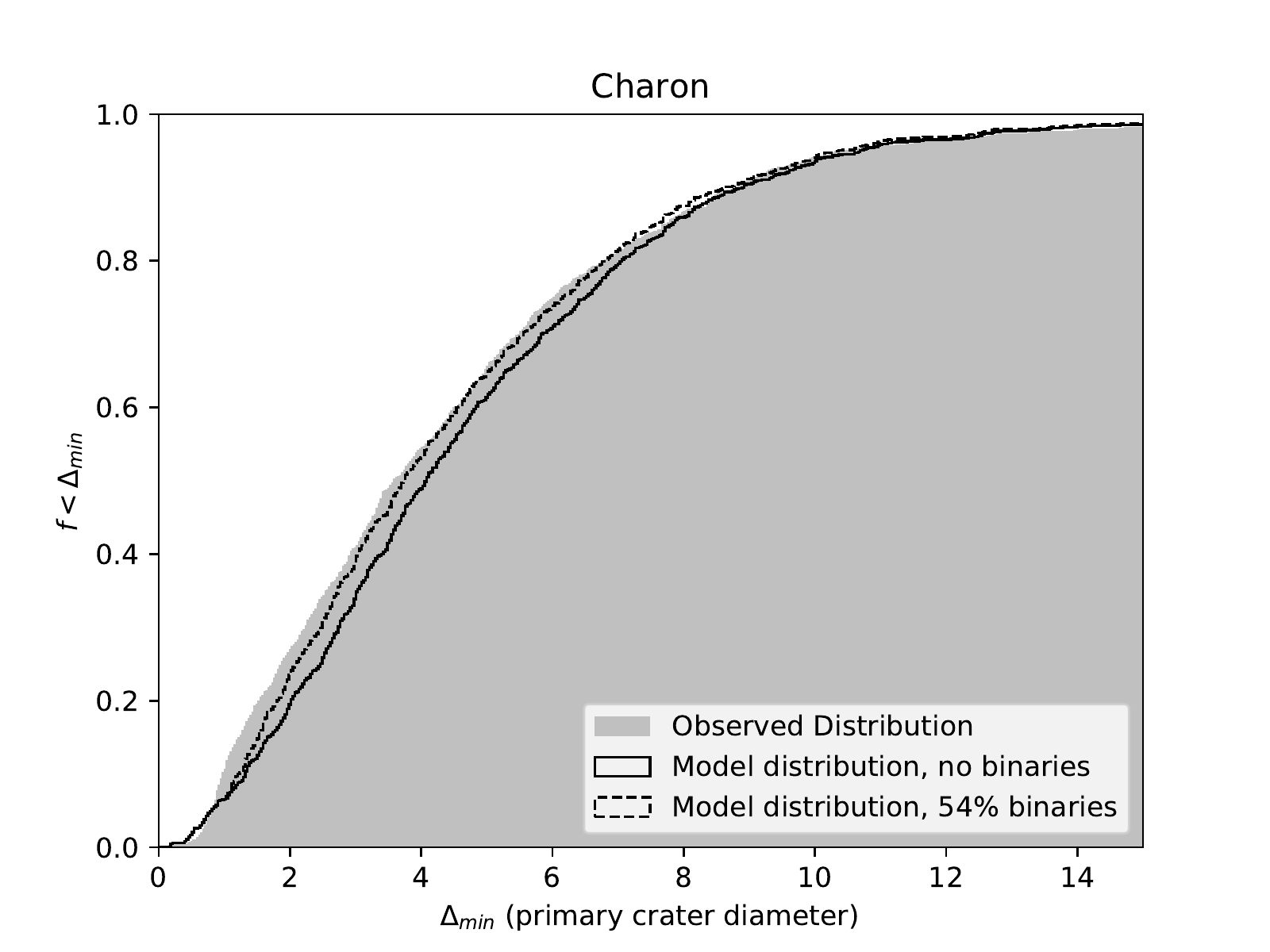}
    \caption{Cumulative distribution of the distances between similarly-sized craters on Pluto and Charon, scaled by primary crater diameter. Instances of model distributions with and without binary populations shown for comparison. A binary-hosting impactor population always generates a $\Delta_{min}$ distribution biased to lower values, which is more consistent with the observed distribution of $\Delta_{min}$.}
    \label{fig:bin_delta}
\end{figure*}

CCKBOs host a large population of widely-separated binary systems (separations reaching up to 10--20\% of the Hill radius; Parker et al. 2011) with near-equal mass components (Petit et al. 2008, Noll et al. 2008, Noll et al. 2020, Parker et al. 2011). These binaries are likely primordial in origin (Petit \& Mousis 2004), sensitive to disruption by any past history of encounters with the giant planets (Parker \& Kavelaars 2010), and sensitive to collisional disruption (Petit \& Mousis 2004, Nesvorn\'{y} 2010, Parker \& Kavelaars 2012). The properties of trans-Neptunian Binaries (TNBs) are powerful tracers of the processes of planetesimal formation and the subsequent dynamical and collisional evolution that the trans-Neptunian populations underwent. 

The apparent occurrence rate of these widely-separated binary systems among the CCKBOs is generally quoted at 20 -- 30\% (e.g., Noll et al. 2008). However, recent work has shown this is strongly influenced by observational selection effects: for example, a binary system is a brighter configuration of a given amount of material than a solitary object, and coupled with the extremely steep mass distribution of the CCKBOs this results in an over-representation of binaries in any flux-limited survey (Benecchi et al. 2018). The \emph{intrinsic} widely-separated binary fraction of the CCKBOs is likely substantially lower, in the range of 15\% -- 20\% for component sizes and separations currently probed by Hubble Space Telescope observations ($d_{p} \gtrsim 30$ km, $a_{m} \gtrsim 3000$ km). Currently, few limits exist to constrain the population of trans-Neptunian binaries at sizes smaller and separations tighter than this.

It is expected that the mutual orbits of many KBO binary systems were modified by a suite of effects referred to as Kozai Cycles with Tidal Friction (KCTF); inclined binary systems will undergo period Kozai cycle oscillations of inclination and eccentricity until tidal friction can halt the oscillations and instead shrink and circularize the binary orbit. A substantial pileup of binary systems resulting from this process may exist at very tight separations -- fractions of a percent of the Hill sphere diameter, well below the resolved limit of HST (Porter \& Grundy 2012). While there is a size dependence on the circularization timescale, if primordial binaries are distributed by some functional scaling of their Hill radius, then the timescale dependence for a rubble pile is $\propto 1/d_{primary}$. Given the speed with which the pileup forms in the simulations of Porter \& Grundy ($\lesssim 10^7$ years), binary systems with components $\sim500$ times smaller than those simulated are likely to have experienced KCTF pileup in the age of the solar system. This implies that binary components down to scale of $d \sim 0.1-0.2$ km, on the order of the size of the smallest impactors recorded by the Pluto and Charon crater records, were subject to modification by KCTF. Thus, a pileup of binary impactors between contact and $0.5\% - 1\%$ of the Hill radius would be a reasonable expectation if the primordial population of binaries had broad inclination, eccentricity, and separation distributions. 

Thus, surfaces within the Pluto system may provide an indirect record of the properties of the binary population at sizes and separations far smaller than have been thoroughly studied by direct means. Binary systems can produce a population of impact craters with spatially-correlated sizes -- that is, craters of similar sizes will tend to fall closer to one another than would be expected under isotropic bombardment by solitary objects. To investigate this prospect, a characteristic distance metric for similarly sized craters must be defined. For each observed crater $i$, the haversine distances to all other observed craters on that target body within the size range $0.7 D_i \leq D_i \leq D_i / 0.7$ is calculated. The smallest distance in this list is divided by $D_i$ to generate $\Delta_{min}$, which is retained as the metric for the best candidate ``doublet crater'' for each observed crater. The distribution of  $\Delta_{min}$ is illustrated in Figure \ref{fig:bin_delta}. To determine if there is evidence for a tightly-separated binary population encoded in the crater sample, a model of the distribution of $\Delta_{min}$ under both the null (no binaries) and alternate (some fraction of binaries) hypotheses must be constructed and tested against the measured distribution of $\Delta_{min}$. Such a model is constructed in the following section.

\bigskip
\textit{2.2.1 Generating Synthetic Crater Doublets}
\bigskip



The process for generating a synthetic crater population that includes binary impactors starts from the same approach as used to generate impactor samples in the previous section. For every sampled impactor, a satellite impactor is generated with probability set by a proposed binary rate. The size of this satellite with respect to the diameter of the primary is sampled from a Rayleigh distribution informed by the brightness ratios of observed widely-separated trans-Neptunian binaries (Johnston, W. R. et al. 2018 and references therein),

\begin{equation}
P(d_{i2}) \propto 
\begin{cases}
0 & d_{i2} < 0.5 d_{i1} \\
\frac{d_{i2}}{(0.2d_{i1})^2} e^{-\frac{d_{i2}^2}{2 (0.2d_{i1})^2)}} & 0.5 d_{i1} \leq d_{i2} \leq d_{i1} \\
0 & d_{i2} > d_{i1}
\end{cases}
\end{equation}

The separation of the two components pre-impact was drawn from a uniform distribution between contact and 0.5\% of the primary's Hill sphere, and their orientation pre-impact was drawn uniformly over the sphere. The impact separation projected onto a target surface was determined by this separation and orientation and the proposed impact angle $\theta$ for a given impactor. The implied crater diameters is generated through the same process as in the preceding section.

Geographic location and regional observational completeness must be accounted for as well. To achieve, one synthetic crater is proposed for every observed crater on each target, and propagate that \emph{real} crater's latitude, longitude, and estimated detection efficiency function from the preceding section to each proposed crater, resampling the diameter from the proposed SFD parameters. Craters created by satellite impactors are retained in the synthetic observed sample using the same detection efficiency function as estimated for the primary, unless they fall within 0.8 $D_i$ of the primary; in this case, it is unlikely that a clear doublet crater would have been generated (e.g., Miljkovi\'{c} et al. 2013). This essentially renders contact binaries (similar to 486958 Arrokoth; Stern et al. 2019) invisible in the crater record.

The haversine distances to all synthetic observed craters within the size range $0.7 D_i \leq D_i \leq D_i / 0.7$ is calculated, and if a proposed satellite produces a crater within this size range and is deemed observed by a trial with the efficiency function, the distance to the crater generated by that satellite is added to this list. The smallest distance in this list is divided by the primary crater diameter to generate $\Delta_{min}$, which is retained as the metric for the best candidate ``doublet crater'' for each crater in the synthetic dataset. Another $10^7$ ABCr trials were conducted, drawing size-frequency distribution parameters and efficiency function parameters from the posterior PDFs determined in the preceding section. In this case, the $D_{ABCr}$ metric is the sum of the KS-test statistics over the observed and proposed $\Delta_{min}$ distributions on Pluto and Charon, truncated at $\Delta_{min} \leq 30$. Simultaneously, a parallel trial was conducted where $\emph{no}$ craters generated by satellites were included in the synthetic proposed sample that defines $\Delta_{min}$. This parallel thread of trials permits the measurement of the Bayes factor in favor of the more complex alternate hypothesis (there is a binary population defined by some population fraction) and the simpler null hypothesis (there is no binary population).

\bigskip
\textit{2.2.2 Binary Population Results}
\bigskip

\begin{figure}[t]
    \centering
    \includegraphics[width=9cm]{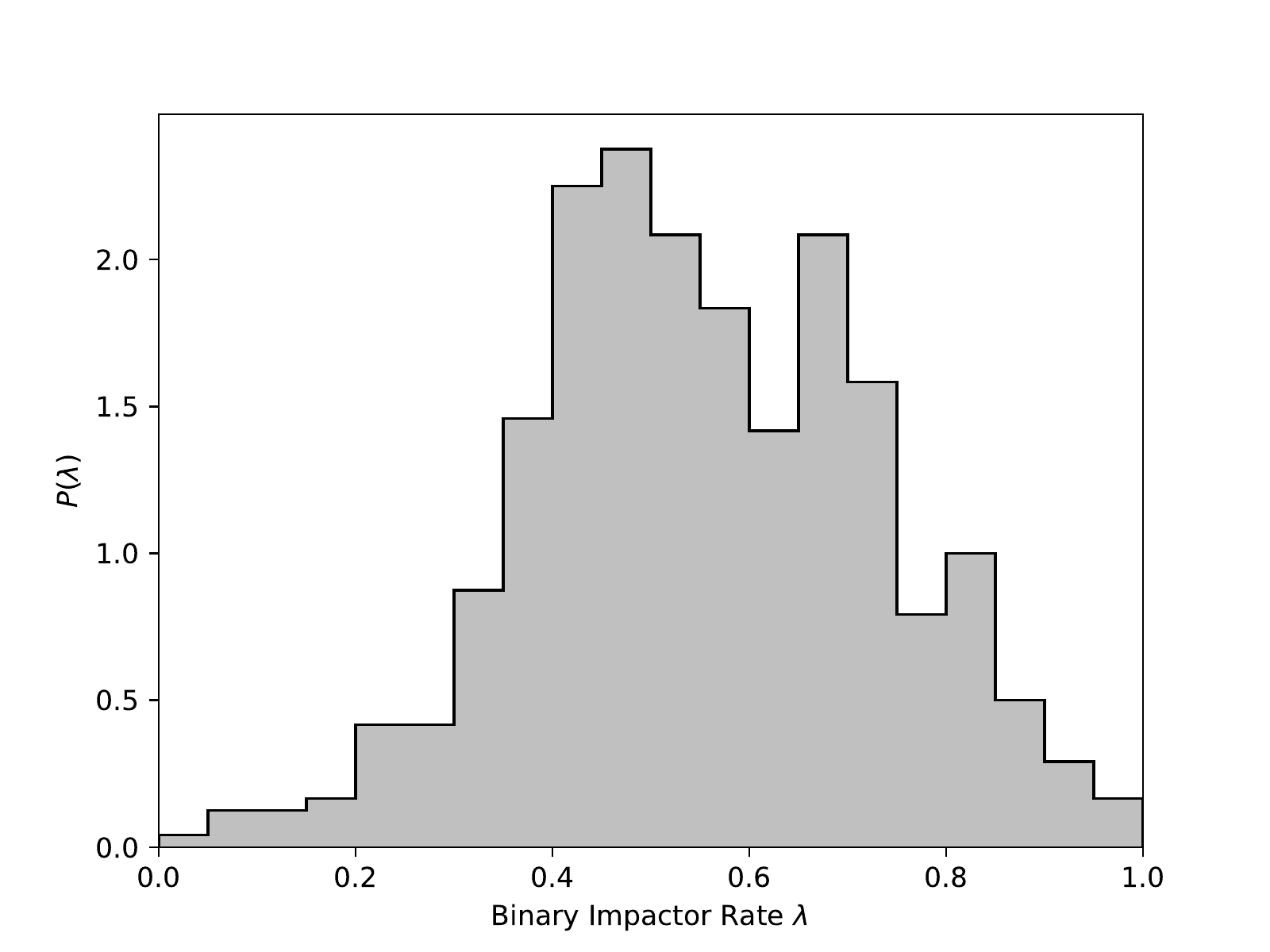}
    \caption{Marginalized posterior PDF of the binary rate parameter $\lambda$ estimated from the separation distribution of similarly-sized craters.}
    \label{fig:bin_frac}
\end{figure}

The acceptance rate of the alternate hypothesis was 22 times that of the null hypothesis -- the Bayes factor is 22:1 in favor of a model with an impactor population that hosts similarly-sized satellites, indicating strong evidence in favor of this hypothesis. The median and 68\% confidence interval of the binary occurrence rate is $\lambda = 0.54^{+0.19}_{-0.16}$; the PDF is illustrated in Figure \ref{fig:bin_frac}. This rate it strongly sensitive to the proposed separation distribution, however -- many of the tightest proposed systems produce single craters, so the uniform distribution proposed can accept high binary fractions with few of them producing craters. Further work to refine the model separation distribution based on KCTF evolutionary models would enable the crater record to better constrain the TNO binary fraction at small sizes; lightcurve and occultation searches for tight binary systems also provide powerful avenues of investigation into this putative population.

\bigskip
\noindent
\textbf{2.3 Implications}
\bigskip

The crater record on the surfaces of Pluto and Charon indicate a very shallow size frequency distribution for small TNOs, and a population of relatively tight binary systems throughout the size range of impactors sampled by the craters ($d \gtrsim 0.1$ km). This shallow slope and binary population are both unlikely to represent the end state of a population that has experienced extensive collisional evolution. While it is conceivable that the CCKBOs are a relatively unmodified primordial population that emerged from an aerodynamically-enhanced collapse of cm-scale ``pebble'' swarms (e.g., Chiang \& Youdin 2010, Nesvorn\'{y} et al. 2010 \& 2019) and their size-frequency distribution and binary population reflect the outcome of that process, the same cannot be said confidently for the populations that were transported to their current orbits from formation locations in the vastly more massive and collisionally active disk closer to the Sun at the time of giant planet migration. If the transported population was extracted from a collisionally evolved source population, why would the SFD slope at small sizes be so shallow? One unexplored possibility is that the transport process itself was size dependent; if smaller objects' collisional interactions tended to drive them out of resonances prematurely compared to their larger counterparts, fewer may have evolved into the stable non-scattering orbits that allowed the larger objects to persist. Future consideration of the the influence of planetesimal size on the effectiveness of posited transportation mechanisms is merited to determine if such processes could help explain the unusual size distribution of small TNOs.

\section{\textbf{MOON ODDS AND ODD MOONS}}

\begin{figure}[t]
    \centering
    \includegraphics[width=9cm]{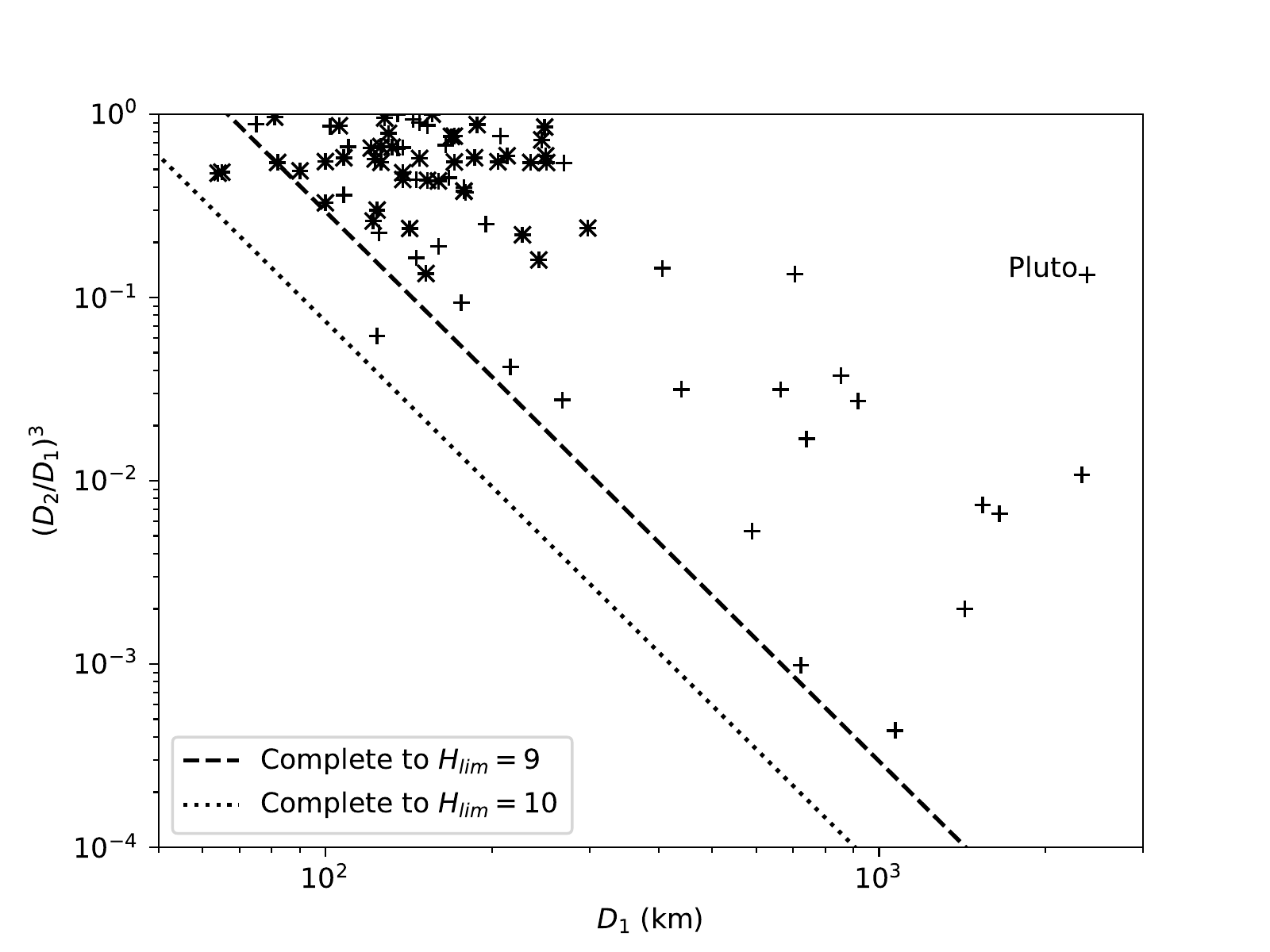}
    \caption{ Diameters and approximate mass ratios of known TNO binary systems. Completeness lines are based on scaling from primary diameter and assuming albedos of 10\% where other estimates are not available. Starred points are CCKBOs. In multi-satellite systems, only the largest satellite is considered. Data from Johnston, W. R. et al. 2018 and references therein.}
    \label{fig:moons}
\end{figure}

\begin{table*}[t]
\centering
\small
\begin{tabular}{ |l|r|c|c|c|c| } 
 \hline
Name/ID     & $H$& Class     & Diam. Measured? & HST search? & Known sats? \\
\hline
\rowcolor{Gray} Eris &          -1.1 & Scattered     & Y & Y & Y (1) \\
\rowcolor{Gray} Pluto &         -0.4 & Resonant      & Y & Y & Y (5) \\
\rowcolor{Gray} Makemake &      -0.1 & Classical     & Y & Y & Y (1) \\
\rowcolor{Gray} Haumea &         0.2 & Resonant      & Y & Y & Y (2) \\
Sedna  &         1.3 & Detached      & Y & Y & N (0) \\
\rowcolor{Gray} Gonggong &       1.6 & Scattered     & Y & Y & Y (1) \\
\rowcolor{Gray} Orcus  &         2.2 & Resonant      & Y & Y & Y (1) \\
\rowcolor{Gray} Quaoar   &       2.4 & Classical     & Y & Y & Y (1) \\
\rowcolor{Gray} 2013 FY$_{27}$ & 3.2 & Scattered     & N & Y & Y (1) \\
                2002 AW$_{197}$& 3.3 & Classical     & Y & Y & N (0) \\
\rowcolor{Gray}  G!k{\'u}n||'h{\`o}md{\'\i}m{\`a} &     3.3 & Scattered     & Y & Y & Y (1) \\
                2002 TX$_{300}$& 3.4 & Haumea Fam.   & Y & Y & N (0) \\
                2014 UZ$_{224}$& 3.4 & Detached      & N & N & ---   \\
                2018 VG$_{18}$ & 3.4 & Detached      & N & N & ---   \\
\rowcolor{Gray} Varda &          3.4 & Classical     & Y & Y & Y (1) \\
                2005 UQ$_{513}$& 3.6 & Classical   & Y & Y & N (0)   \\
                2002 MS$_{4}$  & 3.6 & Classical     & Y & Y & N (0) \\
\rowcolor{Gray} 2003 AZ$_{84}$ & 3.6 & Resonant      & Y & Y & Y (1) \\
                Varuna         & 3.6 & Classical     & Y & Y & N (0) \\  
                2005 QU$_{182}$& 3.6 & Scattered     & Y & N & ---   \\
                Ixion          & 3.6 & Resonant      & Y & Y & N (0) \\
\rowcolor{Gray} 2002 UX$_{25}$ & 3.7 & Classical     & Y & Y & Y (1) \\
                2005 RN$_{43}$ & 3.7 & Classical     & Y & Y & N (0) \\
                2015 RR$_{245}$& 3.8 & Scattered     & N & N & ---   \\
                2002 TC$_{302}$& 3.9 & Resonant      & Y & Y & N (0) \\
                Dziewanna & 3.9 & Scattered     & Y & Y & N (0) \\
  \hline
\end{tabular}
\caption{Current results of satellite searches and diameter measurements (either from radiometry or occultations) among the brightest trans-Neptunian objects. H magnitude estimates extracted from JPL Small-Body Database.}
\label{table:1}
\end{table*}

At first glance, the Pluto system appears strikingly unique in a number of ways: it has the largest primary and the largest secondary of any known dwarf planet system, it has the highest multiplicity known, and it has a very low mass ratio between the primary and secondary. However, these features are each drawn from a distribution across the population of dwarf planets, and in context, a Pluto-like system seems less like an oddball and perhaps more like an inevitability. Figure \ref{fig:moons} illustrates the estimated diameters of TNOs known to host satellites, and the estimated mass ratio of each system. Excluding the CCKBO binary systems, with their small sizes and mass ratios near one, Pluto does not appear so much an outlier in this figure as it does a member of a relatively uniformly-populated distribution. There are smaller primaries with similar mass ratios, and systems with smaller mass ratios with similarly-sized primaries. While the degree of the Pluto system's multiplicity remains unique, Haumea has a complex satellite and ring system, and later sections describe how these two systems may have been even more similar in the past.

Satellites are now known to be ubiquitous for the brightest members of the classical, resonant, and scattered disk populations -- see Table 1. The brightest member of any of these populations that has been searched for satellites with the Hubble Space Telescope (HST) but has no known satellite is 2005 UQ$_{513}$, a classical object with a radiometrically-derived diameter of 498$^{+63}_{-75}$ km (Vilenius et al. 2018). As exemplified by Makemake, some members of this list that have been searched in a single epoch using HST may yet host a detectable satellite (Parker et al. 2016). Among objects in these populations with $H_V > 4$, the current apparent occurrence rate for satellites is 60\%, with 12 of the 20 objects currently characterized by HST known to host satellites. Unlike the smaller objects in the Cold Classical population with similarly-sized primordial binary components (Petit \& Mousis 2004, Nesvorn\'{y} 2011, Parker \& Kavelaars 2012), these large, disimilar-size binary systems are thought to form through processes that occurred after the formation of the parent object. Post-formation giant impacts (McKinnon 1989, Canup 2005, 2011, Leinhardt et al. 2010) are a leading contender for the origin of many of these systems, while capture remains a possibility for others. Regardless, they are generally thought to be the outcome of stochastic events that occurred at some characteristic frequency early in the solar system's history. 

The high extant occurrence rate of satellites around large TNOs thus points to a surprisingly high rate of satellite formation events. Of the 22 systems in Table 1 that are from resonant, scattered, or classical populations, 12 are known to host satellites. Assuming that the extant number of satellites partitions systems into two groups -- one group having one or more satellite formation events per object, and one where zero such events have occurred to date -- the  mean rate parameter for a Poisson distribution of the number of satellite forming events experienced for host objects $H_V > 4$ mag is $\lambda=0.82$. If these events are independent from one another -- that is, one moon-forming event does not influence the probability of another moon-forming event occurring -- then this implies that a substantial fraction of the known large TNOs likely experienced \emph{multiple} moon-forming events in succession, and the current satellite systems are the result of the integrated history of these events for each system. For the same sample, an average of four systems is likely to have experienced at least two moon forming events, and an average of one system will have experienced three or more. For the largest TNOs, histories of multiple satellite formation and disruption events are thus not only possible, but \emph{demanded} by the high extant rate of satellite systems. In this light, it is worthwhile to consider the current complexity of Pluto's satellite system as a potential end-state of one of these multi-epoch formation histories, and examine what can be inferred about the possible past histories of satellite systems that appear less complex or unusual today.

Eris, Makemake, Gonggong, Orcus, and Quaoar all host single known satellites. The largest of these is Eris' satellite Dysnomia, with a diameter of $700 \pm 115$ km estimated from ALMA radiometry (Brown \& Butler 2018). For the most part, little is known about the physical properties of most of these satellites due to the challenges inherent in observing them in proximity to their (typically) much brighter primaries. Gonggong's satellite Xiangliu maintains a substantial eccentricity ($e\sim0.3$, Kiss et al. 2019); such a high eccentricity should have been tidally damped over the age of the solar system unless the tidal factor $Q$ is high and the satellite is small and bright (Kiss et al. 2019), unless it has a recent origin or mechanism for ongoing excitation (such as a second as-yet unseen satellite). 

Another system that invites consideration is that of Haumea. Not only is Haumea host to two satellites and a ring of debris, but it is the largest member of the only known orbital family of TNOs. The largest known member of the Haumea family (other than Haumea itself) is (55636) 2002 TX$_{300}$. With a diameter of 286$\pm10$ km (Elliot et al. 2010), it is comparable in size to Haumea's largest moon Hi'iaka. While larger than the small satellites of Pluto, these objects share the traits of very high albedo surfaces (Weaver et al. 2016), nearly pure water-ice spectra with indications of ammonia compounds (Cook et al. 2018, Barkume et al. 2006). If the Haumea system once bore a resemblance to the Pluto system, with a tight, massive inner binary surrounded by a disk of smaller satellites (though with substantially more mass in this disk than is currently present around Pluto), it is conceivable that instabilities originating within the satellite system could have resulted in the ejection of many of the original satellites at very low velocity with respect to Haumea. Binary systems are very efficient at ejecting material that migrate into an unstable orbital separation regime (Jackson et al. 2018). Should enough orbital energy be removed by repeated ejections of small satellites, an inner binary could be driven to merge into a single, fast-rotating primary (e.g., Levison et al. 2007). Such a scenario would make the Haumea system (and its family) and the Pluto system two outcomes of a relatively similar origin.

If satellite systems growing from circumplanetary debris disks reach the stage of oligarchic growth, the satellites that emerge will have a relatively shallow size-frequency distribution due to the partitioning of their feeding zones. The mid-sized Saturnian satellites display a shallow maximum-likelihood differential SFD slope of 2.4; the inner prograde satellites of Neptune have a very shallow SFD slope of 1.8. New Horizons directly resolved the sizes of the small satellites of Pluto (Weaver et al. 2016), and they show a similar shallow slope of 2.5. Recent measurements have determined that the Haumea family members also have a relatively shallow SFD slope of $\lesssim2.5$ (Pike et al. 2019). These slopes are all shallower than the canonical collisional equilibrium slope of $q=3.5$, and much shallower than typical slopes produced by fragmentation of asteroids as evinced by the size-frequency distribution slopes of young asteroid families ($\bar{q} = 4.46$ for single-slope families younger than 500 Myr; Parker et al. 2008) and also the size distribution slopes produced in graze-and-merge simulations of the origin of Haumea and its family (Leinhardt et al. 2010). The similarity of Pluto's satellites and Haumea's satellites and family (in terms of surface composition, albedo, and size-frequency distributions) may be coincidental, but are certainly suggestive of common origin processes and their relationship merits further study.

\section{\textbf{MORE THAN SKIN DEEP}}

While it is likely to be decades before another trans-Neptunian dwarf planet is revealed at the same level of detail as Pluto and Charon, the properties observed by New Horizons can help guide our understanding of those observations we can currently make of their kin. While one of the chief takeaways of our current exploration of icy worlds in the outer solar system is that they are often more distinct than they are similar, there are still valuable lessons to be learned.

\bigskip
\textit{4.1 Surface composition, atmospheres, and volatile transport}
\bigskip

New Horizons confirmed a long-standing hypothesis regarding the origin of Pluto's high albedo -- that seasonal volatile transport refreshes the deposits of volatiles on its surface, hiding any lag deposits of photochemically-darkened materials (Stern et al. 1988). It is plausible that the surfaces of Makemake and Eris are similarly refreshed by atmospheric processes, resulting in their bright surfaces and their very small rotational lightcurve amplitudes (e.g., Heinze \& de Lahunta 2009, Sicardy et al. 2011, Ortiz et al. 2012, Hofgartner et al. 2019). Models have indicated that these two worlds could host regional or global vapor pressure equilibrium atmospheres of $N_{2}$ and/or $CH_{4}$ at some points in their orbits (e.g., Stern \& Trafton 2008, Young 2015), though searches for atmospheres through stellar occultation measurements have yielded no detectable global atmospheres to date (see Young et al. 2020 for a review). Stellar occultation-based upper limits of surface pressures have been made for Eris (Sicardy et al. 2011), Haumea (Ortiz et al. (2017), Makemake (Ortiz et al. 2012), and Quaoar (Arimatsu et al. 2019b); the $1\sigma$ upper limits range from 1--12$\times10^{-9}$ bar. Ortiz et al. (2012) noted some hints of non-solid-body refraction in the lightcurve of a star occulted by Makemake, and showed that these are consitent with plausible non-global atmospheres, but the data from this event alone is not conclusive.

One of the most striking features on the encounter hemisphere of Pluto is Sputnik Planitia, an impact basin filled by a vast nitrogen-rich ice sheet undergoing active convection (McKinnon et al. 2016). Due to the relatively swift overturn of its convective cells, its surface is extremely young and fresh, showing no impact craters at the limit of the highest resolution imagery from New Horizons (Moore et al. 2016). It has been speculated that the bright, fresh surfaces of other large, volatile-dominated TNOs, such as those of Eris and Makemake, may be maintained in part by similar convective processes at their surfaces (McKinnon et al. 2016, Grundy \& Umurhan 2017). 

Eris has a substantially higher bulk density than Pluto (Sicardy et al. 2011). If this is indicative solely of a difference in rock-to-ice fraction, then Eris can be expected to have substantially greater radiogenic heat production in its interior and thus a higher heat flux at its surface. A larger rock fraction would also imply a greater potential cosmogonic reservoir of nitrogen from which a surface layer of $N_{2}$ could be built up. Both of these factors conspire to create favorable conditions for extensive convective $N_{2}$ deposits on Eris' surface. 

In detail, Eris may contain $\sim1.5 \times 10^{22}$ kg of rock in its interior, suggesting a present-day surface heat flux of order $\sim 5$ mW/m$^{2}$ versus Pluto's $\sim 3$ mW/m$^{2}$.  Following the arguments in McKinnon et al. (2016), the thermal gradient through a surface layer of $N_{2}$ ice is likely to be of order 25 K km$^{-1}$ versus Pluto's 15 K km$^{-1}$. For even the lowest estimates of its surface temperature at aphelion (27 K), this thermal gradient implies that a $N_{2}$ deposits thicker than $\sim1.4$ km would be sufficient to melt $N_{2}$ at its base. Adopting the same cosmogonic nitrogen mass fraction suggested for Pluto by Singer \& Stern (2015), Eris likely contained an initial interior reservoir of $\sim5\times10^{19}$ kg of nitrogen, enough to create a global $N_{2}$ surface layer $\sim3$ km thick, if internal processes ever liberated this material and delivered it to the surface. The ratio of layer thicknesses required to support convection on two different worlds is

\begin{equation}
    \frac{L_1}{L_2} = \exp(T_{i1}^{-1} - T_{i2}^{-1}) \frac{g_2\Delta T_{2}}{g_1\Delta T_{1}},
    \label{eqn:convect}
\end{equation}

where $g$ is the surface gravity of each world, $\Delta T$ is the temperature difference across the layer, and McKinnon et al. (2016) estimate $T_i$ to be $T_0 - \Delta T / 2$ in the sluggish lid regime, and $T_0$ is the basal temperature. Assuming the critical layer thickness for both worlds is lower than the thickness required to generate basal melt, adopting the the thermal gradients above and the relative surface gravities of 0.82 m/s$^2$ and 0.617 m/s$^2$ for Eris and Pluto respectively, and adopting the McKinnon et al. (2016) estimate of $L_{Pluto} \sim 500$ m, Eqn. \ref{eqn:convect} can be solved implicitly to derive $L_{Eris} \sim 390$ m under the same convection regime. Thus, all else being equal, marginally shallower deposits of $N_{2}$ should convect on Eris than do on Pluto.

One caveat here lies in the structure of Pluto's mantle and crust beneath Sputnik Planitia. If the orientation of the basin is due to true polar wander driven by a positive gravity anomaly (Keane et al. 2016, Nimmo et al. 2016), then the lithosphere beneath the basin must be thinned (Johnson et al. 2016). This substantially increases the proximity of the isothermal ocean (see Section 4.3) to the surface underneath Sputnik Planitia, and could act as a thermal conduit and enhance the local heat flux substantially. Thus, Sputnik Planitia could be convective even in a scenario where Pluto's rock component contained a sub-chondritic level of radionuclides; in this case, assuming common composition and absent similar conduits to the interior, Eris may not support nitrogen convection on its surface today, leaving seasonal processing of its surface as the only means of maintaining its high albedo and neutral color.

Atmospheric escape of $N_{2}$ has been put forward as an explanation for the deep methane absorption bands (Licandro et al. 2006), relatively weak evidence for nitrogen in Makemake's spectra (Brown et al. 2007), and lack of a currently-detectable global atmosphere (Ortiz et al. 2012). Given the low current estimates for Makemake's density (1.7 g cm$^{-3}$, based on an equilibrium figure derived from a 7.77 hour spin period\footnote{Recent photometry (Hromakina et al. 2019) suggests that longer rotation periods of $\sim$11.4 hour or $\sim$22.8 hour are likely, and rotational equilibrium figure models will need to be revised for Makemake in future analyses.}; Heinze \& de Lahunta 2009, Rambaux et al. 2017, Parker et al. 2018), Makemake sits at the cusp of where even slow Jeans escape would have depleted its surface of $N_{2}$ over the age of the solar system (Schaller \& Brown 2007, Ortiz et al. 2012, Brown 2012), but still dense enough to have retained $CH_{4}$. Should any substantial deposits of $N_{2}$ persist on the surface of Makemake, convection is challenged by the lower heat flux at its surface ($\sim1.3$ mW m$^{-2}$) driving a lower temperature gradient through any surface deposits ($\sim6.5$ K km$^{-1}$), setting the required layer thickness to support convection to $L_{Makemake}\sim810$ m after accounting for its low surface gravity of 0.36 m/s$^2$. Thus, any $N_{2}$ deposits on Makemake would have to be thicker than Pluto in order to convect, and given Makemake's relative depletion of $N_{2}$, this would seem unlikely. 

Charon's spectrum had revealed ammonia signatures before flyby (Brown \& Calvin 2000, Cook et al. 2007). New Horizons revealed that limited exposures of $NH_{3}$ on Pluto and Charon both appear to be associated with surface modification, including through cryoflows on Pluto (Dalle Ore et al. 2019, Cruikshank et al. 2019) and craters on Charon (Grundy et al. 2016). The presence of absorption features at 2.21 $\mu m$ also indicates ammoniated compounds on the small satellites Nix and Hydra (Cook et al. 2018). As mentioned previously, ammonia compounds have been detected both on Haumea itself and its largest moon Hi'iaka (Barkume et al. 2006). The large TNO Orcus also displays ammonia absorption (Carry et al. 2011). As ammonia is a potent antifreeze agent responsible for lowering the melting point of solid systems of ammonia and water ices, its distribution among the large TNOs is a key factor in understanding the thermal evolution of their interiors and the potential for extant subsurface oceans.

The long-term evolution of Pluto's orbit and obliquity lead to a cycle of 1.5 Myr ``super-seasons'' that may be responsible for controlling the stark albedo and composition contrasts between low and high latitudes through feedback between albedo and voltile transport processes (Earle et al. 2018). Makemake's thermal emission requires two terrains of dissimilar temperature (Lim et al. 2010), which suggests distinct geographic regions of both very low and very high albedo. However, Makemake's extremely small rotational lightcurve amplitude (Heinze \& de Lahunta 2009) and edge-on spin orientation (Parker et al. 2016) require that such dark terrain be distributed extremely longitudinally uniformly. Similar albedo/volatile transport feedback processes could potentially generate latitudinal ``belts'' of dark material on Makemake that, if not interrupted by dissimilar bright terrain (like Sputnik Planitia on Pluto) would be relatively undetectable in the lightcurve. Alternatively, a non-global atmosphere that has frozen out onto the surface near aphelion could produce bright frost-covered terrains around the equator, while older, darker materials are exposed near the poles (Ortiz et al. 2012). The possibility of an upcoming season of Makemake mutual events provides a potential avenue to test for latidudinal variations in albedo on Makemake, potentially isolating the dark terrain and its climatological implications. 

\bigskip
\textit{4.2 Tectonics, Landform Evolution, and Cryovolcanism}
\bigskip

The features attributed to tectonic activity on the surfaces of both Pluto and Charon are predominantly extensional in nature (Moore et al. 2016, Keene et al. 2016, Beyer et al. 2017). The absence of obvious compressional tectonic complexes suggests that both Pluto and Charon have experienced periods of global expansion with no substantial periods of global contraction. Charon's tectonic features, which include graben complexes with enormous vertical relief (up to 5 km), suggest a global areal strain of 1\%, which could be accomplished by freezing a subsurface layer of pure water that was initially $\sim35$ km thick (Beyer et al. 2017). Pluto's extensional features correlate to the location and orientation of Sputnik Planitia in such as a way as to suggest that Pluto experienced true polar wander after the emplacement of the Sputnik Planitia basin, orienting the resultant gravity anomaly along the Pluto-Charon tidal axis (Keene et al. 2016). Very few constraints on similar topographic features exist for large TNOs other than Pluto and Charon; a stellar occultation by the several-hundred kilometer diameter Plutino  2003 AZ$_{84}$ may reveal either a regional depression or a deep chasm seen in profile (Dias-Oliveira et al. 2017).

Unlike Charon, many of Pluto's landforms appear to be sculpted by sublimation processes as part of Pluto's highly active seasonal and ``super-seasonal'' volatile transport cycles (Moore et al. 2016, 2017, 2018, Earle et al. 2018). The striking ``bladed'' terrain on Pluto are regions of uniquely-textured $CH_{4}$-dominated material at low latitude and high elevation. Presumed to be similar to terrestrial water-ice penitentes, these are possibly the result of complex interactions between the thermal profile of Pluto's atmosphere, the seasonally-dependant latitudinal distribution of solar heating, and the interplay of $N_2$ and $CH_4$ precipitation and sublimation (Moore et al. 2018). Pitted terrains with a variety of specific morphologies can be reproduced by sublimation-driven surface evolution (Moore et al. 2017), and the appearance of these pit chains on the surface of the rapidly-refreshed Sputnik Planitia indicates that these landforms are evolving and growing in the current epoch. It is likely that sublimation will be an important process for driving landform evolution on the surfaces of other volatile-dominated TNOs. Non-erosive aeolian processes are also in evidence on Pluto; dune-like features on Sputnik Planitia suggest that small particles of low-density materials have been recently transported by winds near Pluto's surface (Telfer et al. 2018). Whether dynamics within transient atmospheres on other large TNOs are sufficient to similarly mobilize materials at their surfaces is unknown.

A variety of landforms on Pluto -- including extensive dendritic valley networks -- have been interpreted as the result of surface modification by fluid flow (Moore et al. 2016, Howard et al. 2017). A leading contender for the working fluid that produced these is liquid nitrogen. Recent analysis indicates that it is unlikely that conditions have ever existed on Pluto that permitted liquid nitrogen to exist stably on the surface (Bertrand et al. 2018); it is likely that any such $N_2$ flow occurred as basal melt in past subglacial environments. As similar glaciation may be expected on the surfaces of worlds like Eris and Makemake, it is reasonable to expect that similar landforms indicating past fluid flow would be exposed on their surfaces. 

Features on both Pluto and Charon have been identified as potentially cryovolcanic in origin. The smooth terrain of Vulcan Planum on Charon is interpreted as being emplaced by a massive cryoflow onto the surface during an early epoch of global expansion that enabled a subsurface ocean to breach the ice crust in many locations, resulting in an effusive flow of highly viscous material covering a large portion of Charon's surface (Beyer et al. 2019). This material appears to embay older topographic features (including around Charon's striking ``mountains in moats'') and to pile up near its margins south of Serenity Chasma. Vulcan Planitia's crater record is consistent with an age of 4 Gyr (Moore et al. 2016, Singer et al. 2019), though uncertainties in the absolute age calibration may allow ages half that old. If the resurfacing of Vulcan Planum is indicative of an epoch of global expansion, then this age is inconsistent with the origin of this expansion being due to the onset of freezing of a global ocean which is likely to occur much later (e.g., Desch \& Neveu 2017). Recent models of the thermal evolution of Charon indicate that early epochs of global expansion can be generated by the hydrous alteration of silicates in the rocky core (Malamud et al. 2017). 

Located just south of Sputnik Planitia, Pluto's Wright and Piccard Mons are two prominent mound-shaped structures with central depressions. They are both candidates for cryovolcanic features (Moore et al. 2016), and they appear to be relatively young with very few craters found on their surfaces (Singer et al. 2018). Virgil Fossae, just to the west of Sputnik Planitia, is a graben complex with surrounding terrain that shows high concentrations of water ice, a unique red coloring agent, and NH$_{3}$ compounds (Cruikshank et al. 2019). These properties suggest that Virgil Fossae was the source for a regional cryoflow. Wright Mons, Piccard Mons, and Virgil Fossae are all located near the ring-shaped global peak of inferred extensional stress produced by the loading and reorientation of Sputnik Planitia (Keane et al. 2016, Cruikshank et al. 2019). These extensional stresses should act to generate a path of least resistance for any subsurface liquid reservoirs to reach the surface. Given that the most promising candidates for cryovolcanic structures on Pluto currently appear to be limited to these regions of unusually high extensional stresses (though this region also encompasses the best-imaged hemisphere of Pluto, so selection effects may be at play), it is unclear if cryovolcanism on the very largest TNOs would occur without the presence of similar stress-inducing structures like the impact basin underlying Sputnik Planitia. Neveu et al. (2015) explored the role that the exolution of gases from subsurface reservoirs could play in driving ongoing explosive cryovolcanism and found a variety of plausible avenues for such processes, but no active Triton-like plumes were detected in the flyby datasets (Hofgartner et al. 2017).

\bigskip
\textit{4.3 Internal oceans of large TNOs}
\bigskip

\begin{figure}[t]
    \centering
    \includegraphics[width=9cm]{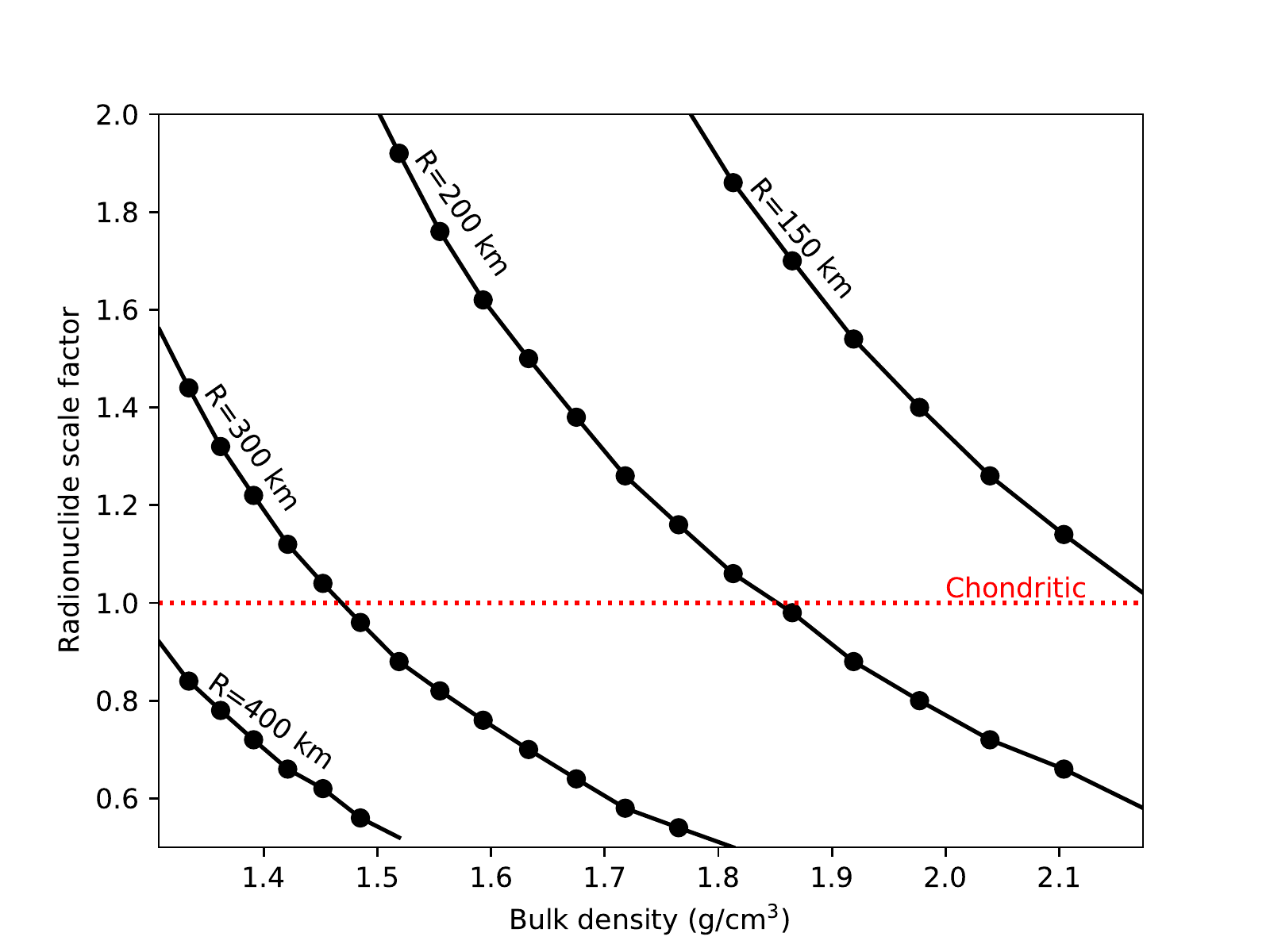}
    \caption{Thresholds for the onset of ammonia dihydrate (ADH) melting in the interior of TNOs containing 1\% ammonia concentration by weight in their non-rock fraction. Bodies accrete cold and begin with uniform distribution of rock and ice throughout their bulk. Nominal specific heat production taken from Desch et al. (2009) shown with red horizontal line.}
    \label{fig:melting}
\end{figure}

Efforts to model the thermal evolution of the interiors of dwarf planets have largely explored systems of water, ammonia, and silicate minerals (Hussmann et al. 2006, McKinnon 2006, Desch et al. 2009, Robuchon \& Nimmo 2011, Rubin et al. 2014, Neveu \& Desch 2015, Desch 2015). The silicate fraction provides a long-lived internal heat source in the form of radiogenic heat, and that heat is transported through the bulk of the body to the surface where it is eventually radiated away to space. Energy is transported radially through conduction and convection. The consensus of modeling efforts before flyby was that (1) the interiors of both Pluto and Charon would be differentiated; (2) if the reference viscosity of their ice mantles is high, then these mantles do not convect and transport energy purely by conduction, generally resulting in the formation of internal oceans; (3) if the reference viscosity of their ice mantles is low, then the mantles convect and efficiently deliver heat from the core to the surface layers, resulting in no ocean formation. Further, models predict that Charon retains an undifferentiated crust of a mix of rock and ice atop a mantle of pure ice (Rubin et al. 2014).

Substantial observational evidence for an internal liquid ocean within Pluto was revealed by New Horizons, including the lack of a fossil bulge, extension-dominated tectonics, and the evidence for true polar wander driven by a gravity anomaly in Sputnik Planitia (Moore et al. 2016, Nimmo et al. 2016, Keane et al. 2016, Johnson et al. 2016) though whether the ocean was transient or persists to this day remains to be determined. Charon's extensional tectonics and cryovolcanic features also support a past internal ocean, though one that has largely or entirely frozen by the current epoch. The extents and survival timescales of similar oceans in trans-Neptunian dwarf planets is a matter of substantial astrobiological interest, as such oceans likely interact directly with silicate cores and provide a potential abode for extraterrestrial life. Subsequent to these discoveries, more detailed interior evolution models were developed for both Pluto and Charon in an effort to explain some of the key observations from flyby -- namely, the relatively similar densities of both worlds, Charon's ancient extensional tectonics, and the gravity anomaly inferred from the orientation of Sputnik Planitia. Both Malamud et al. (2017) and Bierson \& Nimmo (2018) added treatments of porosity removal; Malamud et al. (2017) and Desch \& Neveu (2017) both tracked two rock phases to account for hydrous alteration of silicate minerals. Desch \& Neveu (2017) also included a treatment of suspended rock ``fines'' which produce a more insulating mantle. Kamata et al. (2019) examined the role that clathrate hydrates may play as a thermal insulator between a subsurface ocean and the ice crust above it. All of these additional factors may impact the extent, astrobiological potential, and survival time of interior oceans within large TNOs and the tectonic expression of their evolution on the surfaces of these worlds.

Existing models and comparison to observations within the Pluto system can provide insight into the internal structures of other large TNOs. At a basic level, a conservative estimate of the onset of differentiation can be estimated by determining threshold requirements to heat the interior of the TNO to the melting point of ammonia dihydrate (ADH, $T\sim176$ K); it is likely that differentiation occurs at even lower temperatures than this (e.g., Desch et al. 2015). Figure \ref{fig:melting} illustrates threshold differentiation curves for TNOs of different sizes assuming the Desch et al. (2009) thermophysical model. Relatively small TNOs ($d\lesssim 400$ km) require large rock fractions or high radionuclide enhancement over typical chondritic material to achieve differentiation, but larger TNOs transition to plausible differentiation under nominal conditions. These estimates do not include the influence of early aluminum heating, tidal heating in binary systems, or heating from impacts.

Makemake and Charon have nearly identical bulk densities, and from a thermal evolution standpoint they are relatively similar save for their sizes and an initial epoch of tidal heating for Charon. For bulk ammonia concentrations of 1\% by mass in the initial ice component, Desch et al. (2009) estimate the time to complete ocean freezing of 

\begin{equation}
    t_{freeze} \simeq 4.6 \left( \frac{\bar{\rho}}{1700 \mbox{g cm}^{-3}} \right) \left( \frac{R}{600 \mbox{km}} \right) \mbox{Gyr},
    \label{eqn:melting}
\end{equation}

where $\bar{\rho}$ is the average bulk density of a large TNO and $R$ is its radius. All else being equal, this predicts a freezing timescale of 4.7 Gyr for Charon, and 5.8 Gyr for Makemake. Conversely, Quaoar is very similar in size to Charon (555 km versus 606 km radii; Braga-Ribas et al. 2013), but substantially more dense (2.0 g cm$^{-3}$ versus 1.7 g cm$^{-3}$; Braga-Ribas et al. 2013). The inferred timescale for ocean freezing on Quaoar is also longer than Charon at 5.0 Gyr. While all of these values are model dependent and numerous other models exist that may shorten or extend the lifetime of an internal ocean, they should roughly scale in a similar fashion for these three relatively similar worlds. 

\begin{figure*}[t]
    \centering
    \includegraphics[width=8cm]{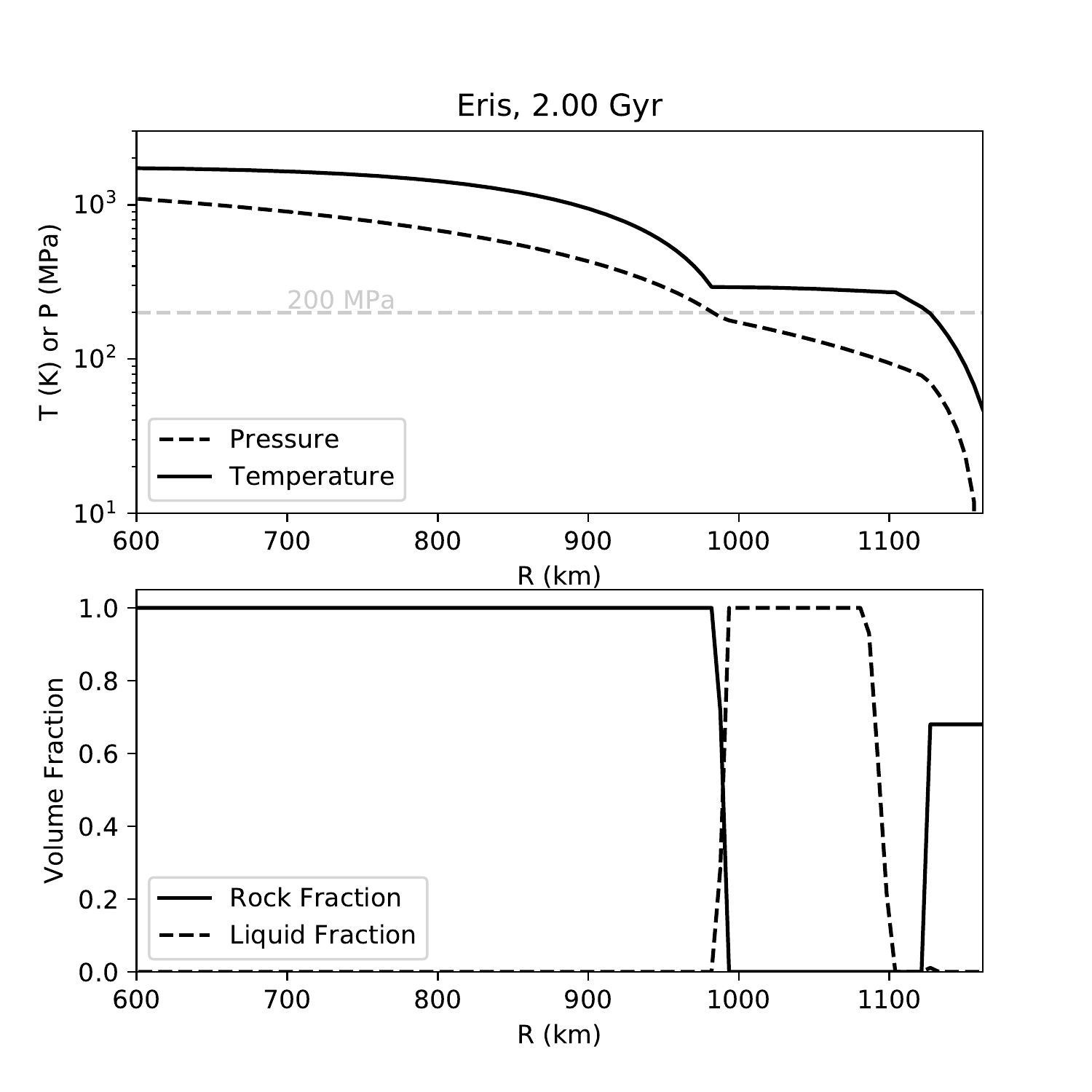}
    \includegraphics[width=8cm]{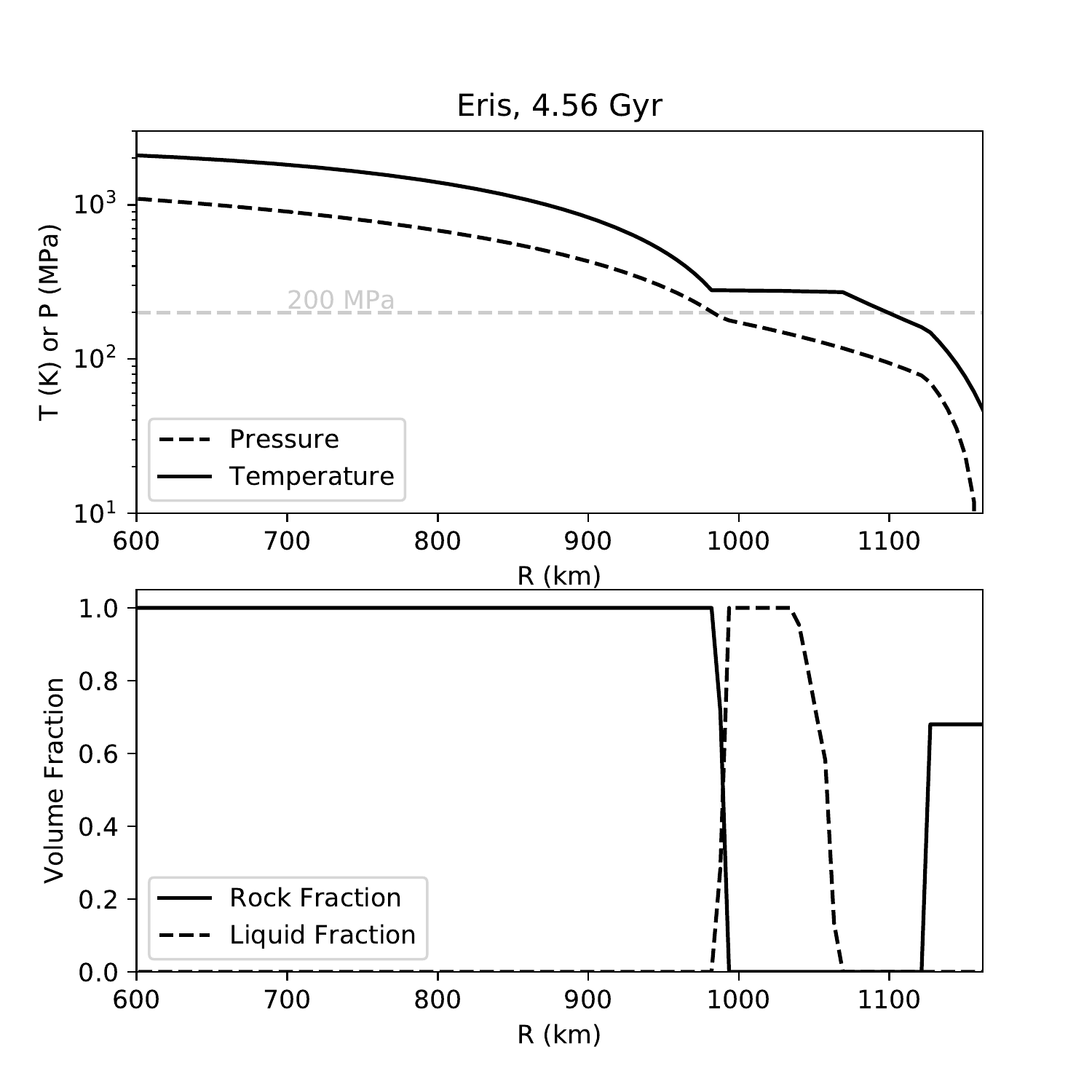}
    
    \caption{Interior properties of Eris at 2 Gyr after formation and at the current epoch, derived from thermophysical models in Desch et al. (2009), assuming only conductive heat transport in the ice shell. Eris rapidly and extensively differentiates, leaving an undifferentiated shell only 41 km thick. Substantial reduction in the thickness of the liquid layer occurs in the last 2 Gyr.}
    \label{fig:eris_interior}
\end{figure*}

Eris and Haumea exist in classes of their own; while they are nearly identical in diameter, Eris' bulk density is substantially higher than Pluto's, and all else being equal this would lead to \emph{lower} pressures at the interface between its ice mantle and rocky core, but substantially higher heat fluxes. Unlike with Pluto, the formation of high density Ice II (rhombohedral) phase of water isn't plausible at its core boundary unless there is a substantial crust of undifferentiated rock-ice mixture to drive up the core-mantle boundary pressure; while the lack of compressional tectonic features on the surface of Pluto indicates that these phases did not form in Pluto's interior (Hammond et al. 2016), fewer potential thermal histories would have produced such effects for Eris. An example of Eris' internal structure's evolution is illustrated in Figure \ref{fig:eris_interior}; the Desch et al. (2009) thermal model predicts a relatively thin undifferentiated layer and the core-mantle boundary remains below the 200 MPa threshold for formation of Ice II. The steeper thermal gradient through the ice crust has greater potential to drive convection in the ice shell, and absent other insulating or viscosity-increasing effects (e.g., clathrate formation, Kamata et al. 2019), the efficient transport of heat from the core to the surface by convection would inhibit ocean formation. If the crust remains conductive, then the scaling from Eqn. \ref{eqn:melting} can be applied to estimate a total freezing time of $\sim13$ Gyr; from 2 Gyr to the present day, approximately 8\% of Eris' volume has undergone a phase transition from liquid water to ice, indicating potential for global expansion and extensional stresses. Haumea's rapid spin and very elongated figure makes 1D thermophysical approximations invalid, and detailed modeling of its interior evolution after acquiring its current physical configuration merits applying a 3D thermophysical scheme. Recent efforts to model the density and figure of Haumea indicate that it is likely differentiated with a dense silicate core and relatively thick ice shell, and that it has achieved a fluid equilibrium figure (Dunham et al. 2019).

As described earlier, the relatively high ratio of $CH_{4}$ to$ N_{2}$ derived from Makemake's NIR spectral properties compared to those of Pluto and Eris suggests that some set of processes have depleted its surface $N_{2}$ relative to these other worlds. While fractionation through atmospheric escape is one possibility that has been explored (e.g., Brown et al. 2007), another possibility is that the internal oceans of Pluto and Eris both host substantial layers of clathrate hydrates (Kamata et al. 2019), which can efficiently sequester $CH_{4}$, but not $N_{2}$. If Makemake's interior is less favorable for the formation of extensive layers of clathrate hydrates, then substantially more of its internal $CH_{4}$ may have been allowed to reach its surface, diluting any $N_{2}$.

\section{\textbf{FUTURE DIRECTIONS AND INFORMED SPECULATION}}

Pluto has revealed a wealth of information about processes that may be endemic to the icy dwarf planets in the outer solar system, or perhaps unique to itself. As the planetary science community continues to tackle the data returned from the Pluto system by New Horizons and collect more remote observations, it will be important to consider which is which. To that end, a robust effort should be undertaken to consolidate the most important components of theory around the origin, structure, and evolution of icy dwarf planets and open them to scrutiny. For example, models of interior evolution of dwarf planets are a field where broad community investment would help drive the field forward, ensuring that the best and broadest expertise can be tapped in the disparate domains needed to achieve high-performance, accessible, accurate, and validated models applicable to the interiors of these worlds. 

Occultation surveys -- both those chasing predicted occultations and those searching for serendipitous events -- provide a clear path forward for resolving the remaining uncertainties in the size-frequency distribution of small TNOs. With sufficient signal-to-noise and multi-cord detections, mapping the shape, radial, and vertical distribution of small TNOs will permit a crude extraction of independent properties of the hot and cold sub-populations. Accurate size-frequency distributions will in theory permit exploration of the surface properties of the Pluto system, including their absolute ages and perhaps their density and porosity. Upcoming programs like TAOS-II (Lehner et al. 2018) are the next step toward these goals, but in the future dedicated space-based serendipitous occultation surveys (Kavelaars et al. 2010, Alcock et al. 2014, Santos-Sanz et al. 2016), above the scintillation noise induced by Earth's atmosphere, will be able to measure to smaller sizes at a greater range of distances and provide higher-quality measurements of each detected object's size and distance. 

The degree of characterization of the properties of other icy dwarf planets will soon be substantially improved by the James Webb Space Telescope (Parker et al. 2016a). Covering the 0.6-28.5 micron spectral range with unprecedented sensitivity and resolution, many fundamental absorption bands for key surface constituents will be readily observable for the first time. JWST's sensitivity to the blue thermal tail and its spatial resolution will permit unprecedented measurements of the thermal properties within satellite systems. Already a large suite of dwarf planet observations are planned to execute as Guaranteed Time Observation programs (GTOs 1191/Stansberry, 1231/Guilbert-Lepoutre, 1254/Parker, 1272/Hines, and 1273/Lunine), with targets including Pluto, Haumea, Makemake, Quaoar, Eris, Sedna, Orcus, and Varuna. These observations will test a variety of observation modes and build an initial framework for determining the most fruitful paths forward for further characterization of these worlds from JWST. Key among these observations is a series to tie the JWST spectral observations to the New Horizons spectral observations of Pluto using a longitudinally-resolved time series of Near Infrared Spectrograph (NIRSpec) observations. Comparing this set of observations with those from New Horizons will provide a critical ``ground truth'' test of what inferences may be drawn from next-generation remote spectral investigations. 

Finally, the population of known dwarf planets continues to grow as several recent surveys have turned up new candidates at very large distances from the sun. As the Large Synoptic Survey Telescope\footnote{\url{https://www.lsst.org/scientists/scibook}} comes online in the early 2020s, we will have a huge leap in survey uniformity and depth over a very large portion of the sky, potentially revealing many as-yet unknown distant dwarf planets. As this population is further revealed and further studied, we can develop a better understanding of just how Pluto fits into the picture. As we go forward with our efforts to discover more of its far-flung kin, Pluto's singular story can only deepen our understanding of the yet more distant worlds waiting to be found.

\section{\textbf{SUMMARY AND CONCLUSIONS}}

By acting as a witness plate recording eons of bombardment by small TNOs, the Pluto system has provided us with a powerful means for estimating the TNO size-frequency distribution at sizes far smaller than existing direct-detection surveys. Folding in existing direct-detection constraints at larger sizes and using a likelihood-free inference method indicates that the size distribution of small TNOs breaks from a slope of $q_1 \simeq 5.15$ (for hot populations) or $q_1 \simeq 8.50$ (for cold populations) at diameters larger than 100 km to a population-average slope of $q_2 \simeq 2.90$ down to sizes of $d \sim 2.8$ km, where it breaks again to an even shallower slope of $q_3 \simeq 1.85$ down to sizes as small as $\simeq 0.1$ km, limited by the current resolution of the crater record. The exact location of the break location between $q_3$ and $q_2$ is poorly constrained, perhaps suggesting that the SFD behavior in this regime is more complex than modeled. While this SFD does not support some models of TNO growth motivated by populations inferred from occultation detections (e.g., Schlichting et al. 2013), it is not wildly inconsistent with the occultations themselves.

Further, craters appear to have a size-spatial correlation; that is, craters of similar sizes prefer to be closer to one another than they should be if they were randomly distributed on the surface. This effect persists after accounting for correlated discovery sensitivity and other effects. The hypothesis that this size-spatial correlation is due to the presence of a population of modestly-separated binary impactors is tested with a similar likelihood-free inference approach. This analysis indicates strong evidence in favor of this hypothesis and conclude that there may be a pileup of tidally-modified non-contact binaries with separations in the range of a few to tens of primary radii. The crater record suggests this population exists down to very small sizes. 

Pluto's remarkable satellite system is also examined in the context of the ubiquity of satellite systems among the largest TNOs. This ubiquity indicates that a number of these large TNOs must have exerienced multiple moon-forming events early in their history, and the Pluto system may be the outcome of one of these multiple-event histories. The similarities of the Pluto and Haumea systems point to a novel concept for the origin of Haumea's satellites, its spin and figure, and its family -- that it is an alternate end state for a system that once looked similar to that of Pluto. 

The expression of processes that have been active on the surfaces of Pluto and Charon provide insights into those processes that may have shaped other large TNOs. Conditions on Eris are favorable for hosting extensive convective nitrogen ice deposits, providing another potential avenue to explain its fresh, bright, volatile-dominated surface in spite of photochemical processes that should rapidly alter and darken it. Makemake's smaller size and density make it less favorable for these processes. Thermophysical models indicate that internal oceans may yet survive within many of the largest TNOs, even if one does not persist within Charon.

\quad

\textit{\textbf{Acknowledgements.}} This work was supported in part by the New Horizons mission. Development of the size distribution inference tools was supported by STScI grant AR 14309. The author thanks H. Levison, S. Protopapa, L. Young, B. Bottke, K. Nowicki, D. Nesvorn\'{y}, K. Kratter, and S. Robbins for insightful discussions regarding many aspects of this document. The author would also like to thank the three chapter reviewers -- B. Holler, K. Singer, and one anonymous reviewer -- for their close reads of the document and productive criticisms.


\nocite{*}

\bigskip



\bibliographystyle{apalike}
\bibliography{references}

\end{document}